\documentclass[preprint,12pt]{elsarticle2}

\usepackage{graphicx}

\usepackage{amssymb}

\usepackage[subfigure]{tocloft}
\usepackage{subfigure}
\usepackage{float}
\usepackage{natbib}
\usepackage{epsf,ngerman}
\usepackage{latexsym}
\usepackage{graphics}
\usepackage{geometry}
\usepackage[T1]{fontenc}
\usepackage{longtable}
\usepackage{amsmath}
\usepackage{textcomp}
\usepackage[nooneline]{caption}
\usepackage{multirow}
\usepackage{array}

\usepackage[ngerman]{babel}
\usepackage[latin1]{inputenc}
\usepackage{natbib}
\usepackage{amssymb}
\usepackage{times} 
\usepackage[pdftex]{color}
\usepackage{placeins} 
\usepackage{colortbl}
\usepackage{chngcntr}
\usepackage{float}
\usepackage{eurosym}
\usepackage{pdfpages}
\usepackage{rotating}

\begin{document}

\begin{frontmatter}



\title{Everything counts! - Warum kleine Gemeinden die Gewinner der Zensuserhebung 2011 sind
}

\address[SFTH]{Adresse: Prof. Dr. Bj\"orn Christensen, Fachhochschule Kiel, Fachbereich Wirtschaft, Institut f\"ur Statistik und Operations Research, Professur f\"ur Statistik und Mathematik, Sokratesplatz 2, D-24149 Kiel, Germany, Tel:  +49 431 210 3560, E-Mail: bjoern.christensen@fh-kiel.de; \\
Dr. S\"oren Christensen, Universit\"at Hamburg, Fachbereich Mathematik, Bereich Mathematische Statistik und Stochastische Prozesse, Bundesstraße 55, 20146 Hamburg, Germany, Tel. ++49 40 42838 4934, E-Mail: soeren.christensen@uni-hamburg.de; \\
Dr. Tim Hoppe, Amt f\"ur Statistik Magdeburg, 39104 Magdeburg, Germany, \\Tel: +49 391 540-2808 (Fax: -2807), E-Mail: tim.hoppe@stat.magdeburg.de; \\
Michael Spandel, Amt f\"ur Statistk Magdeburg, 39104 Magdeburg, Germany, \\Tel: +49 391 540-2390 (Fax: -2807), E-Mail: michael.spandel@stat.magdeburg
.de}


\author{Bj\"orn Christensen, S\"oren Christensen, Tim Hoppe und Michael Spandel}

\address{Fachhochschule Kiel, Christian-Albrechts-Universit\"at Kiel und Amt f\"ur Statistik Magdeburg}

\begin{abstract}
Mit der Durchf\"uhrung des Zensus 2011 ist die Bundesrepublik Deutschland ihrer Verpflichtung aus der EU-Verordnung 763/2008 vom 9. Juli 2008, eine  Volksz\"ahlung nach festgelegten Kriterien durchzuf\"uhren, nachgekommen. Im Gegensatz zu fr\"uheren Volksz\"ahlungen (1987 BRD und 1981 DDR) wurde beim Zensus 2011 nicht auf eine\linebreak Vollerhebung zur\"uckgegriffen, sondern ein registergest\"utztes Verfahren verwendet. F\"ur die Gemeinden stehen nun, durch die Ver\"offentlichung der Ergebnisse am 31.05.2013, unter anderem die amtlichen Bev\"olkerungszahlen zur Verf\"ugung. Der Großteil der Gemeinden sieht sich elementaren Verringerungen innerhalb ihres Einwohnerbestands gegen\"uber. Dieses Papier zeigt auf, dass Gemeinden unterhalb von 10.000 Einwohner signifikant geringere relative Verluste in den Einwohnerzahlen durch den Zensus 2011 im Vergleich zu Gemeinden mit mehr als 10.000 Einwohner aufweisen.
  \end{abstract}


\end{frontmatter}
\pagebreak

\interfootnotelinepenalty 10000

\section{Motivation}\label{K1}
Die Anzahl der Einwohner spielt f\"ur jede Gemeinde eine \"uberaus wichtige Rolle. Sei es nun relevant f\"ur die Stadtplanung oder f\"ur ordnungsrelvante Aspekte. Gleichzeitig bestimmt sich \"uber die Einwohnerzahl aber auch die Zuteilung von finanziellen Mitteln \"uber den L\"anderfinanzausgleich, die Einteilung von Wahlbezirken oder die Entlohnung der B\"urgermeister. Deshalb gilt, dass schon ein einzelner fehlender Einwohner einen starken Effekt nach sich ziehen kann. Aus diesem Grund haben die St\"adte und Gemeinden gespannt auf die erste Ver\"offentlichung zu den Zensusergebnissen gewartet. Wie bedeutend der Zensus 2011 sowohl f\"ur die kommunale Statistik, als auch f\"ur die Wissenschaft ist, zeigen Egeler et. al (2012) auf. So weisen sie daraufhin, dass der Zensus 2011 als Kombination von Nutzung administrativer Daten und Befragung methodisch neue Wege bestreitet. Dabei hat sich die Zusammenarbeit von amtlicher Statistik und Wissenschaft insbesondere bei der Entwicklung des Stichprobendesigns und des Hochrechnungsverfahrens bew\"ahrt. Der Zensus 2011 wurde neben des vom Bundesministerium des Inneren einberufenen Expertenteams zudem von Mitarbeitern der Universit\"at Trier und des GESIS - Leibniz-Institut f\"ur Sozialwissenschaften wissenschaftlich betreut.\\

Zensen stellen keine Errungenschaften der letzten beiden Jahrhunderte dar.\footnote{F\"ur eine historische W\"urdigung sei an dieser Stelle auf Grohmann (2000) verwiesen.} In den vorangegangenen Volksz\"ahlungen in der Bundesrepublik wurde bisher auf die klassische Methode der Vollerhebung zur\"uckgegriffen.\footnote{Lediglich bei der Volksz\"ahlung 1970 wurde ein Teil der Fragen  auf einer 10-\%igen Stichprobe gest\"utzt (Grohmann (2009)).} So beschreibt Grohmann (2009), dass das bisherige Programm der Volksz\"ahlung 1950 zu einer Volks-, Berufs-, Geb\"aude-, Wohnungs- und Arbeitsst\"attenz\"ahlung erweitert wurde, was man auch in den Jahren 1961, 1970 und 1987 beibehielt. Dabei konnten mit Hilfe von Frageb\"ogen demografische und sozio-\"okonomische Daten abgefragt. Die Verteilung und das Einsammeln der Frageb\"ogen erfolgte durch besonders verpflichtete "Z\"ahler".\\

Die Volksz\"ahlung im Jahr 1987 stand unter keinem guten Stern. Angefangen bei Volksz\"ahlungsboykotts, \"uber die Argumentation des Entstehens eines gl\"asernen B\"urgers aufgrund der Totalerhebung von Daten, Protestbewegungen, bis zu einer Klage vor dem Bundesverfassungsgericht (siehe Scheuch et al. (1989) und Grohmann (2009)). Gegenstand der Kritik war dabei die nicht sicherzustellende Anonymit\"at der Einzeldaten unter Ber\"ucksichtigung der elektronischen Datenverarbeitung. Im Urteil des Bundesverfassungsgerichts vom 15.12.1983 wurde, bis auf den Melderegisterabgleich und zwei Vorschriften \"uber die Weiterleitung personenbezogener Daten ohne ausreichende Anonymisierung, die Verfassungsbeschwerde zur\"uckgewiesen. Allerdings verlangte das Verfassungsgericht vom Gesetzgeber eine erneute Methodendiskussion f\"ur kommende Volksz\"ahlungen (Grohmann (2009)).\\

Der Wandel in der angewandten Methodik begann  Anfang der 90er Jahre. So erarbeitete im Jahr 1995 eine Arbeitsgruppe der amtlichen Statistik einen Bericht, der 12 verschiedene Verfahrensans\"atze einander gegen\"uberstellt (Grohmann (2009)). Ein Jahr sp\"ater einigte sich die Innenministerkonferenz darauf, die bisher verwendete Volksz\"ahlung einzustellen und stattdessen einen registergest\"utzten Zensus zu verwenden. Eine von dieser Konferenz beauftragte Arbeitsgruppe entwickelte daraufhin zwei Modelle. Im Jahr 2001 wurde dann mit der Erprobung eines registergest\"utzten Zensus, dem sogenannten Zensustest begonnen. Ziel dabei war es, neben dem \"ubergang auf die Registerauswertung, auch den Anteil an prim\"arstatistischer Erhebung unter dem Aspekt der damit verbundenen hohen Kosten zu reduzieren (Braun (2004)). Dabei kam man zu dem Ergebnis, dass ein registergest\"utzter Zensus in Deutschland grunds\"atzlich m\"oglich ist. \\

Mit dem Datum 29.08.2006 stand der Beschluss fest, dass sich die Bundesrepublik an den kommenden EU-weiten Volksz\"ahlungen beteiligen wird. Durch das Zensusvorbereitungsgesetz vom 13.12.2007 wurde der Aufbau des Anschriften- und Geb\"auderegisters angeordnet. Inhaltlich sollten f\"ur jedes Wohngeb\"aude Merkmale wie Postleitzahl, Ort, Straße und Hausnummer, Personenzahl sowie weitere Daten erfasst werden (Grohmann (2009)). Nur etwa ein Jahr sp\"ater wurde das Gesetz zur Anordnung des Zensus 2011 beschlossen. Wichtiges Element dabei war die Haushaltsbefragung auf Stichprobenbasis, die mit einer Auskunftspflicht der Befragten versehen wurde. Die Methodik verfolgte zwei Ziele. Zum einen die Karteileichen und Fehlbest\"ande innerhalb der Melderegister zu identifizieren und somit gleichzeitig die neue amtliche Einwohnerzahl zu ermitteln. Auf der anderen Seite sollten durch die Stichprobenerhebung Erhebungsmerkmale gesammelt werden, die in der EU-Verodnung vom 02.09.2008 aufgef\"uhrt wurden. Elementar war dabei, dass die Haushaltsbefragung auf Stichprobenbasis nur f\"ur Gemeinden mit 10.000 Einwohnern und mehr Anwendung finden sollte. Grohmann (2009) weist in diesem Zusammenhang auf Folgendes hin: "Die durch die Beschr\"ankung der Korrektur-Stichprobe auf die Gemeinden mit mindestens 10.000 Einwohnern verbleibenden Fehler in den kleineren Gemeinden werden in Kauf genommen, auch wenn sie m\"oglicherweise sp\"ater Proteste von negativ betroffenen Gemeinden ausl\"osen k\"onnen". \\

Mit dem Stichtag 09.05.2011 wurde der Zensus 2011 durchgef\"uhrt, knapp 2 Jahre sp\"ater liegen die daraus resultierenden Ergebnisse vor. Dieses Papier wertet die Ergebnisse der amtlichen Einwohnerzahl aus dem Zensus 2011 insbesondere vor dem Hintergrund der unterschiedlich verwendeten Methoden f\"ur Gemeinden unter 10.000 Einwohner und ab 10.000 Einwohner aus. Abschnitt~\ref{K2} beschreibt die im Zensus 2011 verwendeten Methoden zur Berechnung der amtlichen Einwohnerzahl. Im Abschnitt~\ref{K3} wird die verwendete Datengrundlage aus den Ergebnissen des Zensus 2011 aufgezeigt. Die sich daraus ergebenden Resultate sind unter Abschnitt~\ref{K4} und Abschnitt~\ref{K5} zu finden. Das Papier schließt mit der Conclusion.

\section{Die Methode zur Berechnung der amtlichen Einwohnerzahl}\label{K2}

F\"ur die Berechnung der amtlichen Einwohnerzahl muss das Gesamtkonzept betrachtet werden. Dies beginnt mit der Festlegung der Stichprobe. Hinsichtlich der Stichprobe wurde f\"ur den Zensus 2011 auf eine geschichtete Stichprobe zur\"uckgegriffen, wobei insgesamt 8 Schichten verwendet wurden, die sich abh\"angig von der gemeldeten Anzahl der Einwohner an einer Anschrift ergaben. Als Qualit\"atsmerkmal wurde dabei der relative Root Mean Square Error ausgew\"ahlt, der f\"ur den gesch\"atzten Totalwert von Merkmalsauspr\"agungen nicht gr\"oßer als 15\% sein d\"urfte. Eine Analyse der Stichprobenqualit\"at ist allerdings nicht Untersuchungsgegenstand dieses Papiers. Aus diesem Grund wird f\"ur tiefgehende Beschreibungen des Stichprobendesigns auf die Arbeit von Berg und Bihler (2011) verwiesen.\\  

Bei der Ermittlung der amtlichen Einwohnerzahl wurden, in Abh\"angigkeit der Gr\"oße der Gemeinde, unterschiedliche Vorgehensweisen verwendet, wobei die Bundesregierung, das Statistische Bundesamt sowie die Landes\"amter auf Grundlage des Zensustests 2001 davon ausgehen, dass beide Wege zu einer vergleichbaren Genauigkeit der Ergebnisse f\"uhren (Sinner-Bartels (2013)). Dabei stellt eine Einwohnerzahl von 10.000 die Trennlinie zwischen den Gemeinden dar, und bestimmt somit zu welcher Verfahrensgruppe zur Bestimmung der amtlichen Einwohnerzahl eine Gemeinde geh\"ort.\\

In Gemeinden mit mindestens 10.000 Einwohnern wurde mit Hilfe der Haushaltebefragung auf Stichprobenbasis die statistische Bereinigung der Fehler in den Melderegistern durchgef\"uhrt. Um eine h\"ohere Genauigkeit der Ergebnisse zu erlangen, wurden die St\"adte und Gemeinden in Deutschland in 4 sogenannte Sampling Point Typen eingeteilt, wobei die Klassifizierung wie folgt vorgenommen wurde (siehe dazu auch M\"unnich et al. (2012)):

\begin{itemize}
	\item Typ 0 (SDT): Stadtteile ab 200.000 Einwohner (EW) aus Gemeinden mit mindestens 400.000 EW
	\item Typ 1 (GEM): Gemeinden mit mindestens 10.000 EW, sofern sie nicht zum Typ 0 geh\"oren
	\item Typ 2 (VBG): Kleine Gemeinden (unter 10.000 EW) innerhalb eines Gemeindeverbands beziehungsweise
einer Verbandsgemeinde werden zusammengefasst, sofern sie in der Summe mindestens
10.000 EW betragen
	\item Typ 3 (KRS): Zusammenfassung aller Gemeinden eines Kreises, die bis dahin noch keinem Typ
zugeordnet wurden 
\end{itemize}

Die Stichprobenanschriften wurden dann im Verlauf des Zensus von den Erhebungsbeauftragten aufgesucht und die Existenz der dort gemeldeten Personen gekl\"art sowie durch die Befragung die weiteren ben\"otigten Merkmale erfasst. Durch einen Abgleich mit dem beim Statistischen Bundesamt gef\"uhrten Referenzdatenbestand wurden die Karteileichen und Fehlbest\"ande festgelegt, worauf sich die Hochrechnung der Stichprobenergebnisse unter Ber\"ucksichtigung der Karteileichen und Fehlbest\"ande anschloss (Sinner-Bartels (2013)). Bei der Hochrechnung wurde auf ein Verfahren der Wissenschaftler M\"unnich und Gabler zur\"uckgegriffen. M\"unnich et al. (2012) \"uberpr\"ufen mehrere Sch\"atzmethoden auf ihre Eignung f\"ur die Berechnung der amtlichen Bev\"olkerungszahl. Es stellte sich heraus, dass der verallgemeinerte Regressionssch\"atzer (GREG) als geeignet zur Hochrechnung der amtlichen Einwohnerzahlen erscheint. Gleichung~\ref{M1} stellt die Struktur des GREG Sch\"atzers dar.\footnote{Das in Gleichung~\ref{M1} befindliche Indizes $d$ steht f\"ur die Domain. Nach M\"unnich et al. (2012) ist also davon auszugehen, dass f\"ur die Sch\"atzung mehrere Gemeinden einer Domain verwendet werden sollen. Nach Aussage des Statistischen Bundesamtes auf den Infoveranstaltungen zum Zensus 2011 f\"ur die Kommunen, wurden die Sch\"atzungen allerdings seperat f\"ur die jeweiligen Gemeinden durchgef\"uhrt. Dies wurde auch nochmals telefonisch best\"atigt.}

\begin{equation}\label{M1}
\hat{\tau}^{GREG}_{Y,d}=\sum_{i\in S_d}{w_{i,d}y_{i,d}}+\left(\sum^{N_d}_{i=1}{x_{i,d}}-\sum_{i\in S_d}{w_{i,d}\cdot x_{i,d}}\right)\hat{\beta}
\end{equation}

mit

\begin{equation}\label{M1}
\hat{\beta}=\left(\sum_{i\in S}{w_ix_ix_i'}\right)^{-1}\sum_{i\in S}{w_ix_iy_i}
\end{equation}

Dabei sind $x_i$ die Hilfsinformationen der i-ten Anschrift, die auch vektoriell vorliegen k\"onnen. $\beta$ ist
die L\"osung der KQ-Sch\"atzung des Regressionskoeffizienten im linearen Regressionsmodell bezogen auf die gesamte Stichprobe $S$ (M\"unnich et al. (2012)) .\\

F\"ur Gemeinden mit weniger als 10.000 Einwohner war die Haushaltsbefragung auf Stichprobenbasis dagegen nicht geeignet. Sinner-Bartels (2013) schreibt dazu: "Aufgrund der Erfahrungen aus dem Zensustest ist in Gemeinden mit weniger als 10.000 Einwohnern eine Stichprobe zur Korrektur der \"uber- und Untererfassungen in den Angaben aus den Melderegistern nicht mehr effizient, weil viel zu hohe Auswahls\"atze erforderlich w\"aren". Aus diesem Grund wurde f\"ur diese Gemeinden die Befragung zur Kl\"arung von Unstimmigkeiten durchgef\"uhrt, wenn sich an Anschriften entsprechend unplausible Konstellationen von Informationen aus den Melderegistern und den Erhebungsbefunden aus der Geb\"aude- und Wohnraumz\"ahlung ergaben.\footnote{Die Geb\"aude- und Wohnraumz\"ahlung wurde unabh\"angig von der Einwohnerzahl in allen St\"adten und Gemeinden durchgef\"uhrt. Diese Z\"ahlung stellt eine Vollerhebung dar.} Sinner-Bartels (2013) weist in diesem Zusammenhang daraufhin, dass mit diesem Verfahren davon auszugehen ist, dass durch die Bereinigung der Karteileichen und Fehlbest\"ande die Registerfehlerrate der kleinen Gemeinden \"ahnlich gut abgesenkt werden kann, wie durch die hochgerechneten Ergebnisse der Haushaltsstichprobe in den gr\"oßeren Gemeinden.\\

Festzuhalten ist, dass das Statistische Bundesamt und die Landes\"amter sich das Ziel gesetzt hatten, obwohl zwei unterschiedliche Methoden f\"ur Gemeinden unter und ab 10.000 Einwohner verwendet werden, einen identischen Qualit\"atsstandard zu erf\"ullen. Damit sollte auch der Aspekt einhergehen, dass keine der verwendeten Methoden zu besseren Ergebnissen in der relativen Ver\"anderung zwischen den alten amtlichen Einwohnerzahlen und den aus dem Zensus 2011 resultierenden amtlichen Einwohnerzahlen f\"uhrt. \\  

\section{Daten}\label{K3}

F\"ur die folgenden Betrachtungen wurde auf eine Ver\"offentlichung des Statistischen Bundesamtes zur\"uckgegriffen. Im Gemeindeverzeichnis-Informationssystem auf der Seite www.destatis.de wird eine Datei bereitgestellt, die f\"ur die Gemeinden in Deutschland nach Bev\"olkerung am 31.12.2011 auf Grundlage des Zensus 2011 und fr\"uherer Z\"ahlungen einen Vergleich aufstellt.\footnote{https://www.destatis.de/DE/ZahlenFakten/LaenderRegionen/Regionales/Gemeindeverzeichnis/\linebreak Administrativ/AdministrativeUebersicht.html} In dieser Datei werden die Zahlen der amtlichen Bev\"olkerung nach dem Zensus 2011 sowie der fr\"uheren Berechnung gegen\"ubergestellt. Gleichzeitig werden die absoluten und relativen Ver\"anderungen der Einwohnerzahlen aufgezeigt. Zus\"atzlich enth\"alt diese Datei den Gemeindenamen, den Gemeindeschl\"ussel sowie das Bundesland dem die Gemeinde zuzuordnen ist. F\"ur unsere Analyse betrachten wir die Ergebnisse f\"ur Gemeinden ab 1.000 Einwohner.\footnote{Der Ausschluss von Gemeinden mit weniger als 1.000 Einwohnern ergibt sich aus den z.T. sehr hohen relativen Ver\"anderungen f\"ur diese. Es sei allerdings darauf hingewiesen, dass die Regressionsergebnisse aus Tabelle 1 im folgenden Kapitel 4 in allen wesentlichen Ergebnissen (Insignifikanz f\"ur die Einwohnerzahl und signifikanter negativer Einfluss des Methoden-Dummys) best\"atigt werden k\"onnen, wenn auch diese Kommunen in die Analysen eingeschlossen werden.}  Entsprechend ergibt sich f\"ur das Bundesgebiet eine Menge von 7.148 Beobachtungen. 

\section{Ergebnisse}\label{K4}
\begin{figure}[h]
\caption{Mittlere relative Ver\"anderung zwischen amtlicher Bev\"olkerungszahl aus dem Zensus 2011 und der bisherigen amtlichen Einwohnerzahl}\label{abb1}
\begin{center}
 \includegraphics[width=14cm,height=10cm]{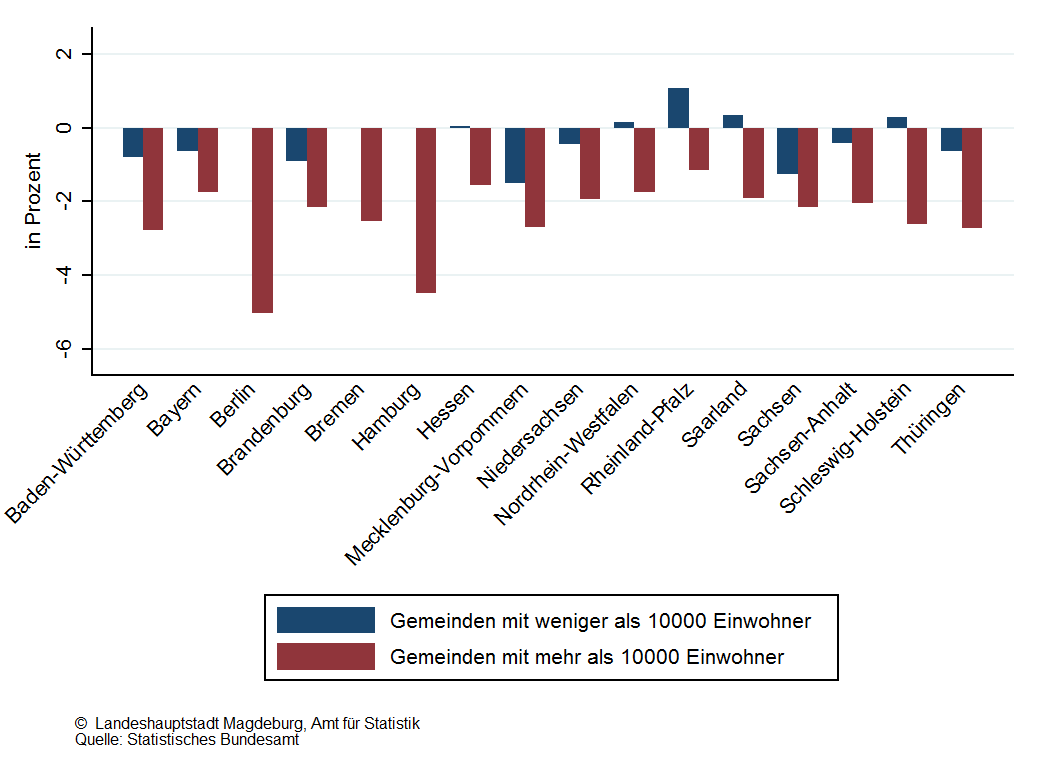}
 \end{center}
\end{figure}

Auf der Grundlage der ver\"offentlichten Gemeindedaten durch das Statistische Bundesamt sollen in diesem Abschnitt einige Analysen durchgef\"uhrt werden. Ziel dabei ist es zu kl\"aren, ob die unterschiedlichen Methoden zur Berechnung der amtlichen Bev\"olkerungszahl aus dem Zensus 2011 f\"ur Gemeinden unter 10.000 Einwohner und ab 10.000 Einwohner keine signifikanten Unterschiede in der relativen Ver\"anderung der amtlichen Einwohnerzahlen zur Folge hat. Abbildung~\ref{abb1} zeigt die mittlere relative Ver\"anderung zwischen amtlicher Bev\"olkerungszahl aus dem Zensus 2011 und der bisherigen amtlichen  Einwohnerzahl differenziert nach den einzelnen Bundesl\"andern auf.\footnote{Die Bundesl\"ander Berlin, Hamburg und Bremen fehlen in der Abbildung~\ref{abb1}, da hier ein Vergleich zwischen Gemeinden unter 10.000 Einwohnern und Gemeinden \"uber 10.000 Einwohner nicht m\"oglich ist.}\\

Aus Abbildung~\ref{abb1} geht hervor, dass f\"ur den Großteil der Bundesl\"ander, unabh\"angig der Methodengrenze von 10.000 Einwohner, eine negative mittlere relative Ver\"anderung vorliegt. Lediglich in den Bundesl\"andern Hessen, Nordrhein-Westfalen, Rheinland-Pfalz, Saarland sowie Schleswig-Holstein ist der Mittelwert der relativen Ver\"anderung positiv. Dieses Ergebnis gilt allerdings nur f\"ur Gemeinden unter 10.000 Einwohner. Wie in Abbildung~\ref{abb1} zudem zu erkennen ist, f\"allt die mittlere relative Ver\"anderung der Gemeinden unter 10.000 Einwohner, verglichen zu der mittleren relativen Ver\"anderung f\"ur Gemeinden ab 10.000 Einwohner, in allen Bundesl\"andern positiver aus. Da die beiden Gruppen unabh\"angige Stichproben darstellen, wurden die relativen Ver\"anderungen der Gemeinden unter 10.000 Einwohner mit den relativen Ver\"anderungen der Gemeinden ab 10.000 Einwohner mit Hilfe des Mann-Whitney-U-Test auf Signifikanz \"uberpr\"uft. Als Ergebnis ergibt sich hierf\"ur, dass Gemeinden mit unter 10.000 Einwohner signifikant bessere relative Ver\"anderungen aufweisen (zweiseitiger Mann-Whitney-U-Test, $p<0.001$).\\

\begin{figure}[h]
\caption{Balkendiagramm relative Ver\"anderung nach Gemeindegr\"oßen }\label{abb2}
\begin{center}
 \includegraphics[width=12cm,height=9cm]{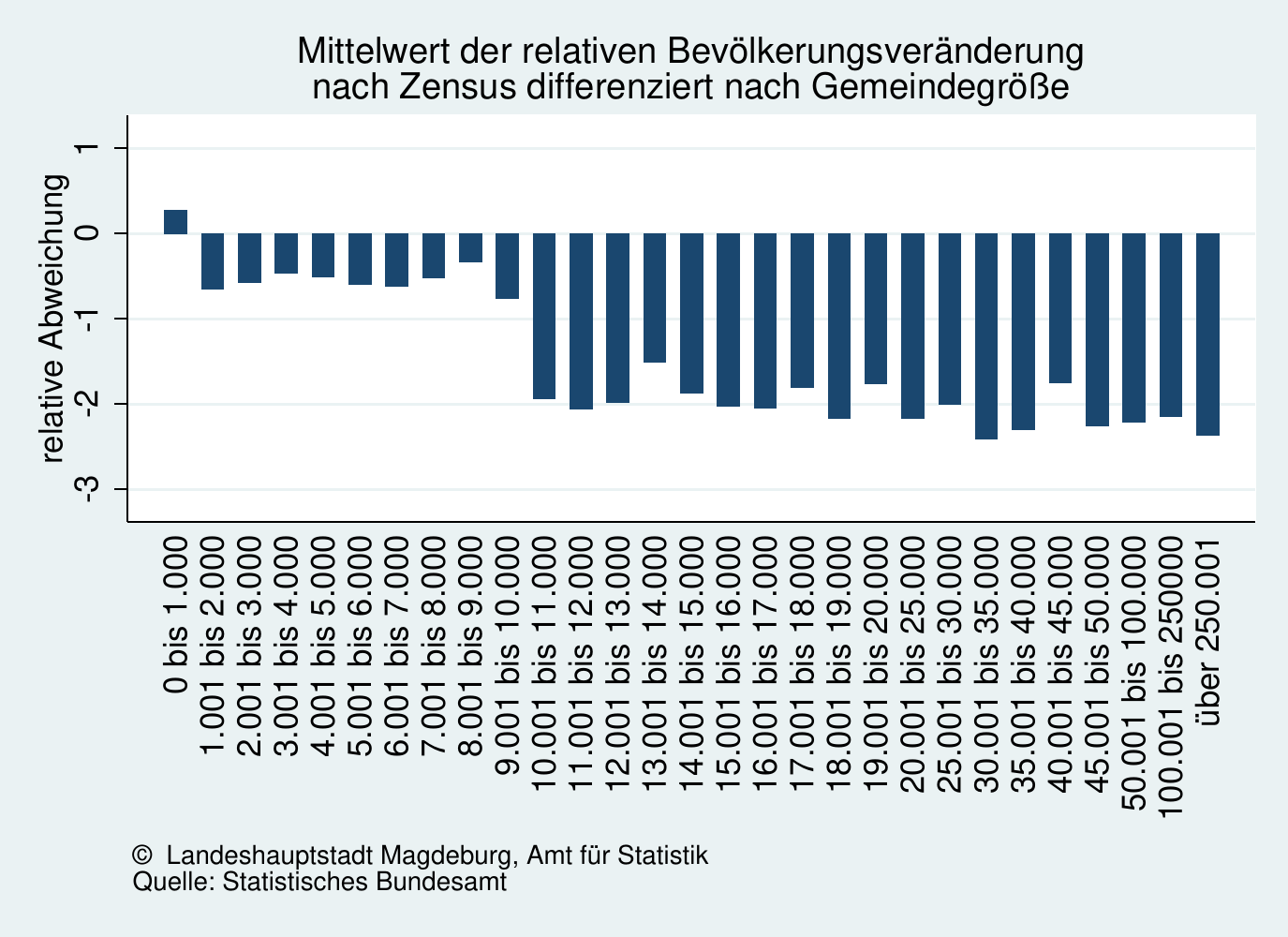}
 \end{center}
\end{figure}

Dabei stellt sich nun die Frage, ob die Unterschiede aufgrund der differenziert verwendeten Methoden resultieren, oder gr\"oßere Melderegister einfach qualitativ weniger gut gef\"uhrt sind bzw. schwerer zu f\"uhren sind. Dazu sei auf Abbildung~\ref{abb2} verwiesen, in der ersichtlich wird, wie stark sich der Methodenwechsel bei 10.000 Einwohnern auf die relativen Ver\"anderungen f\"ur Gesamtdeutschland auswirkt. Es ist deutlich ein Strukturbruch zu erkennen, der exakt an der Grenze von Gemeinden mit weniger bzw. ab 10.000 Einwohnern auftritt. Dar\"uber hinaus ist zu erkennen, dass -- wenn \"uberhaupt -- \"uber den Strukturbruch hinaus nur einen geringen Gr\"oßeneinfluss der Einwohnerzahl auf die relative Ver\"anderung vorliegt.\\

\begin{figure}[h]
\caption{Scatterplot bisherige amtliche Einwohnerzahl zu den relativen Ver\"anderungen Sachsen-Anhalt }\label{abb3}
\begin{center}
 \includegraphics[width=12cm,height=9cm]{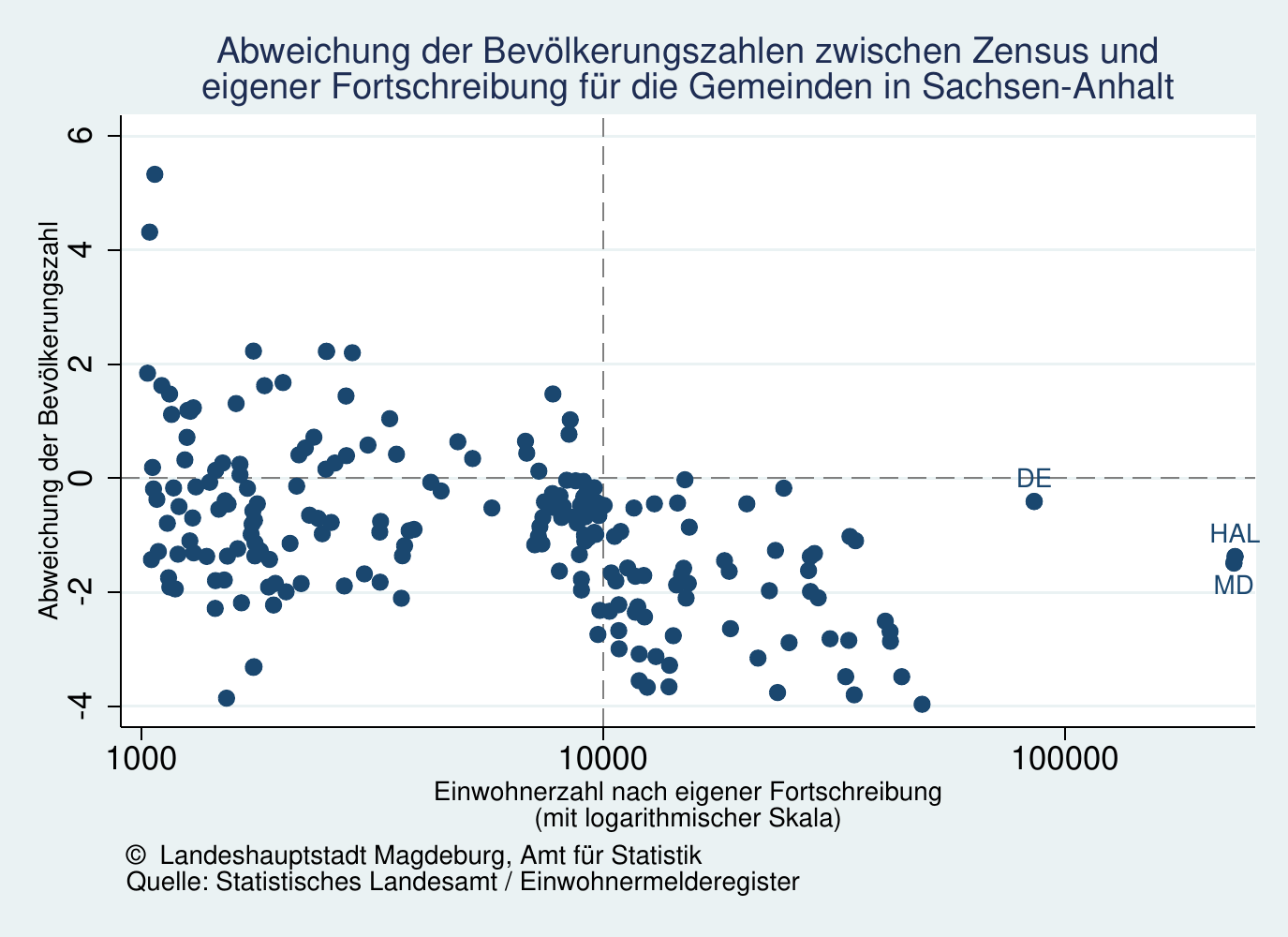}
 \end{center}
\end{figure}

Dieser Strukturbruch wird im Folgenden genauer am Beispiel Sachsen-Anhalts untersucht. Daf\"ur zeigt die Abbildung~\ref{abb3} einen Scatterplot f\"ur die bisherige amtliche Einwohnerzahl zu den relativen Ver\"anderungen f\"ur das Land Sachsen-Anhalt.\footnote{F\"ur die \"ubrigen Bundesl\"ander mit Ausnahme von Berlin, Hamburg und Bremen wurden diese Grafiken ebenfalls erstellt.}  \\

Um eine bessere Darstellung zu gew\"ahrleisten, wurde die x-Achse f\"ur die bisherigen amtlichen Einwohnerzahlen logarithmiert. Weiterhin wurde in der Abbildung~\ref{abb3} die Methodengrenze eingezeichnet. Bei der Betrachtung der Abbildung~\ref{abb3} stellt man fest, dass sich die relativen Ver\"anderungen f\"ur Gemeinden unter 10.000 Einwohner in Sachsen-Anhalt sowohl im positiven Bereich als auch im negativen Bereich befinden. Das heißt, auch in Sachsen-Anhalt gibt es Gemeinden, die nach dem Zensus mehr Einwohner im Vergleich zur alten Einwohnerfortschreibung aufweisen. Des Weiteren erkennt man aber auch, dass alle Gemeinden ab 10.000 Einwohner Verluste in der Einwohnerzahl durch den Zensus 2011 verzeichnen m\"ussen. Die relativen Ver\"anderungen liegen hier im negativen Bereich. Die drei großen St\"adte in Sachsen-Anhalt Magdeburg, Halle(Saale) und Dessau-Roßlau sind besonders markiert. Zwar verliert die Stadt Dessau-Roßlau lediglich moderat. Magdeburg und Halle weisen dagegen sehr identische relative Ver\"anderungswerte auf, die im mittleren negativen Bereich liegen. Betrachtet man den Scatterplot genauer, so scheint es, dass die Punktwolke leicht fallend mit den bisherigen amtlichen Einwohnerzahlen verl\"auft. Dabei scheint dieser Trend jedoch recht schwach ausgepr\"agt. \\

Um die Frage zu beantworten, ob nun die zwei unterschiedlichen Methoden oder die Qualit\"atsverluste der Melderegister bei steigender Einwohnerzahl die Begr\"undung f\"ur die schlechteren relativen Ver\"anderungen bei Gemeinden ab 10.000 Einwohner sind, sollen im Folgenden Regressionsanalysen durchgef\"uhrt werden. Hierf\"ur werden die Daten des Statistischen Bundesamtes mit einer multiplen Regression analysiert. Dabei wird die relative Ver\"anderung der Einwohnerzahl in allen Regressionsberechnungen als endogene Variable verwendet. F\"ur das Land Sachsen-Anhalt wird die Regression beispielhaft auf Bundesl\"anderebene durchgef\"uhrt.\footnote{Diese Regressionsanalyse wurde ebenfalls f\"ur die \"ubrigen Bundesl\"ander mit Ausnahme von Berlin, Hamburg und Bremen berechnet.} Als exogene Variable wird zum einen die bisherige amtliche Einwohnerzahl verwendet. Ein signifikant negativer Effekt w\"urde in diesem Fall bedeuten, dass jeder zus\"atzliche Einwohner die relative Ver\"anderung negativ beeinflusst. Dies ist gleichzusetzen damit, dass die Qualit\"at der Melderegister bei h\"oheren Einwohnerzahlen eher schlechter einzusch\"atzen ist und zu viele Karteileichen beinhaltet. Zus\"atzlich wird in das Regressionsmodell eine Dummy-Variable eingef\"ugt, die Auskunft dar\"uber gibt, ob das Stichprobenverfahren angewendet wurde. Diese Dummy-Variable nimmt nur f\"ur St\"adte ab 10.000 Einwohner den Wert "Eins" an. Das identische Regressionsmodell wird zudem f\"ur die Werte f\"ur alle Gemeinden in Deutschland angewendet.\\

Um analysieren zu k\"onnen inwieweit bundeslandspezifische Einfl\"usse existieren, erweitern wir das Regressionsmodell um $(n-1)$ Dummyvariablen f\"ur die einzelnen Bundesl\"ander, wobei $n$ die Zahl der Bundesl\"ander ist. Tabelle~\ref{Tab1} zeigt die Ergebnisse f\"ur die drei durchgef\"uhrten Regressionen.\\

\begin{table}
\caption{Ergebnisse der Regressionen}\label{Tab1}
\begin{center}
\scalebox{0.80}{%
\begin{tabular}{lccc} \hline
 & (1) & (2) & (3) \\
VARIABLEN & Relative Ver\"anderung  & Relative Ver\"anderung  & Relative Ver\"anderung  \\ 
& Sachsen-Anhalt & Deutschland & Deutschland \\ \hline
Methode & -1.637*** & -1.410*** & -1.505*** \\
& (0.215) & (0.0715) & (0.0799) \\
Einwohnerzahl & 1.31e-06 & -7.71e-07 & -5.62e-07 \\
 & (3.98e-06) & (5.03e-07) & (8.50e-07) \\
Schleswig-Holstein &  &  & 0.813*** \\
 &  &  & (0.180) \\
Hamburg &  &  & -0.955 \\
 &  &  & (2.828) \\
Niedersachsen &  &  & 0.619*** \\
 &  &  & (0.159) \\
Bremen &  &  & 0.170 \\
 &  &  & (1.717) \\
Nordrhein-Westfalen &  &  & 0.852*** \\
 &  &  & (0.189) \\
Hessen &  &  & 1.008*** \\
 &  &  & (0.178) \\
Rheinland-Pfalz &  &  & 1.308*** \\
 &  &  & (0.161) \\
Baden-W\"urttemberg &  &  & -0.00434 \\
 &  &  & (0.153) \\
Bayern &  &  & 0.376*** \\
 &  &  & (0.144) \\
Saarland &  &  & 0.815** \\
 &  &  & (0.360) \\
Berlin &  &  & -0.537 \\
 &  &  & (3.800) \\
Brandenburg &  &  & 0.0337 \\
 &  &  & (0.197) \\
Mecklenburg-Vorpommern &  &  & -0.0921 \\
 &  &  & (0.205) \\
Sachsen &  &  & -0.134 \\
 &  &  & (0.174) \\
Sachsen-Anhalt &  &  & 0.557*** \\
 &  &  & (0.215) \\
Constant & -0.437*** & -0.576*** & -1.013*** \\
 & (0.107) & (0.0326) & (0.133) \\
Beobachtungen & 201 & 7,148 & 7,148 \\
$R^2$ & 0.262 & 0.057 & 0.087 \\
 Adj. $R^2$ & 0.254 & 0.0568 & 0.0849 \\ \hline
\multicolumn{4}{c}{Standardfehler in Klammern} \\
\multicolumn{4}{c}{ *** p$<$0,01, ** p$<$0,05, * p$<$0,1} \\
\end{tabular}}

\end{center}
\end{table}

Wie die Koeffizienten f\"ur die Einwohner in der Tabelle~\ref{Tab1} deutlich zeigen, liegt hier kein elementarer Effekt vor. F\"ur Sachsen-Anhalt ergibt sich ein positiver Effekt, f\"ur die Bundesrepublik Deutschland in beiden Modellen dagegen ein negativer Effekt. Festzuhalten ist, dass diese Effekte so marginal sind, dass sie keinen Erkl\"arungsgehalt f\"ur die Fragestellung aufweisen, warum die relative Ver\"anderung in Gemeinden ab 10.000 Einwohner so deutlich schlechter ausf\"allt als in den Gemeinden mit Einwohnerzahlen unterhalb von 10.000 Einwohner. Somit kann ein m\"ogliches Argument, dass gr\"oßere St\"adte qualitativ schlechtere Register haben, nicht best\"atigt werden. Unterst\"utzt wird dies dadurch, dass in keinem der drei Modelle eine Signifikanz f\"ur die exogene Variable Einwohner vorliegt.\\

Ein anderes Bild ergibt sich f\"ur die Variable Methode, die als Dummy-Variable f\"ur die unterschiedlich verwendeten Methoden zwischen Gemeinden unter und ab 10.000 Einwohner steht. F\"ur die Regressionsanalyse f\"ur das Bundesland Sachsen-Anhalt ergibt sich ein Wert von -1,637. Das bedeutet, dass die relative Ver\"anderung in Gemeinden in Sachsen-Anhalt mit mehr als 10.000 Einwohner, im Vergleich zu Gemeinden mit weniger als 10.000 Einwohner, um 1,637 Prozentpunkte schlechter ausf\"allt.\footnote{Diese Regressionsanalyse wurde f\"ur alle weiteren Bundesl\"ander mit Ausnahme von Berlin, Hamburg und Bremen ebenfalls durchgef\"uhrt. Dabei konnte festgestellt werden, dass die Koeffizienten f\"ur die exogene Variable Methode durchweg negativ sind. Des Weiteren sind diese Koeffizienten hoch signifikant auf einem Niveau von $p<0.01$.} F\"ur Deutschland ergibt die Regressionsanalyse ohne Bundesl\"ander-Dummy-Variablen einen Koeffizienten von -1,410. Verwendet man f\"ur die Bundesl\"ander Dummy-Variablen, so f\"allt der Wert des Koeffizienten f\"ur die Variable Methode auf -1,505. Beachtenswert dabei ist, dass sich f\"ur alle drei Modelle hoch signifikante Ergebnisse finden lassen. So sind die Koeffizienten signifikant auf einem Niveau von $p<0,01$. Aufgrund dieser Ergebnisse l\"asst sich deutlich aufzeigen, dass durch die Methode der Haushaltsbefragung auf Stichprobenbasis, Gemeinden ab 10.000 Einwohner schlechtere Ergebnisse im Zensus 2011 erzielen. Im Vergleich dazu profitieren Gemeinden unter 10.000 Einwohner, da hier auf eine alternative Methodik zur\"uckgegriffen wurde. \\


Im dritten Regressionsmodell mit Dummy-Variablen f\"ur die Bundesl\"ander best\"atigen sich die Erkenntnisse, die bereits vom Statistischen Bundesamt pr\"asentiert wurden. So weisen Gemeinden aus Rheinland-Pfalz zum Basisbundesland Th\"uringen die gr\"oßten positiven Effekte auf. Dieser Effekt ist  signifikant auf einem Niveau von $p<0,01$. Weiterhin weisen auch Hessen oder Nordrhein-Westfalen positive Koeffizienten f\"ur ihre Dummy-Variablen auf, die zudem hoch signifikant sind. Zu den Verlierern geh\"oren dagegen die Bundesl\"ander Berlin, Hamburg, Sachsen oder Baden-W\"urttemberg. Die hierf\"ur vorliegenden Koeffizienten sind allerdings nicht statistisch signifikant.\\        

Auch ein nicht-linearer Einfluss der Einwohnerzahl auf die relative Ver\"anderung ist grunds\"atzlich nicht auszuschließen. Aus diesem Grunde wurden folgende Transformationen der Einwohnerzahl in der Regressionsanalyse ber\"ucksichtigt: quadratische Effekte, kubische Effekte, ein exponentieller Einfluss sowie ein inverser Einfluss. In allen vier F\"allen bleiben die Einfl\"usse der Einwohnerzahl in den Regressionen insignifikant. Zus\"atzlich \"andert sich das Bestimmtheitsmaß nicht und der Koeffizient der Variable Methode liegt jeweils zwischen -1,39 und -1,48.  Der Einfluss des Methoden-Dummy ist dabei weiterhin in allen F\"allen signifikant mit p < 0,01. Um weitere m\"ogliche Gr\"oßeneffekte der Einwohnerzahl zu untersuchen, wurden die Regressionen auch unter Ausschluss von St\"adten mit mehr als 1 Million Einwohnern wiederholt, welches keine relevanten \"anderungen bewirkt.

\section{Erweiterte Analysen}

Die in dem vorangegangenen Kapitel gemachten Analysen k\"onnen unter Umst\"anden differenzierte Wirkungszusammenh\"ange darstellen. Aus diesem Grund werden im Folgenden die Analysen um zus\"atzliche Hypothesen und den dazugeh\"origen statistischen Methoden erweitert. Ziel dieses Kapitels ist es, die bereits vorgestellten Ergebnisse auf ihre Robustheit hin zu \"uberpr\"ufen.\\

\begin{figure}[h]

\caption{Scatterplotbetrachtung f\"ur festgelegte Intervalle}\label{abb4}
\begin{center}
\subfigure[Intervall von 5.000 bis 15.000 Einwohner\label{abb4a}]{\includegraphics[width=7cm,height=7.5cm]{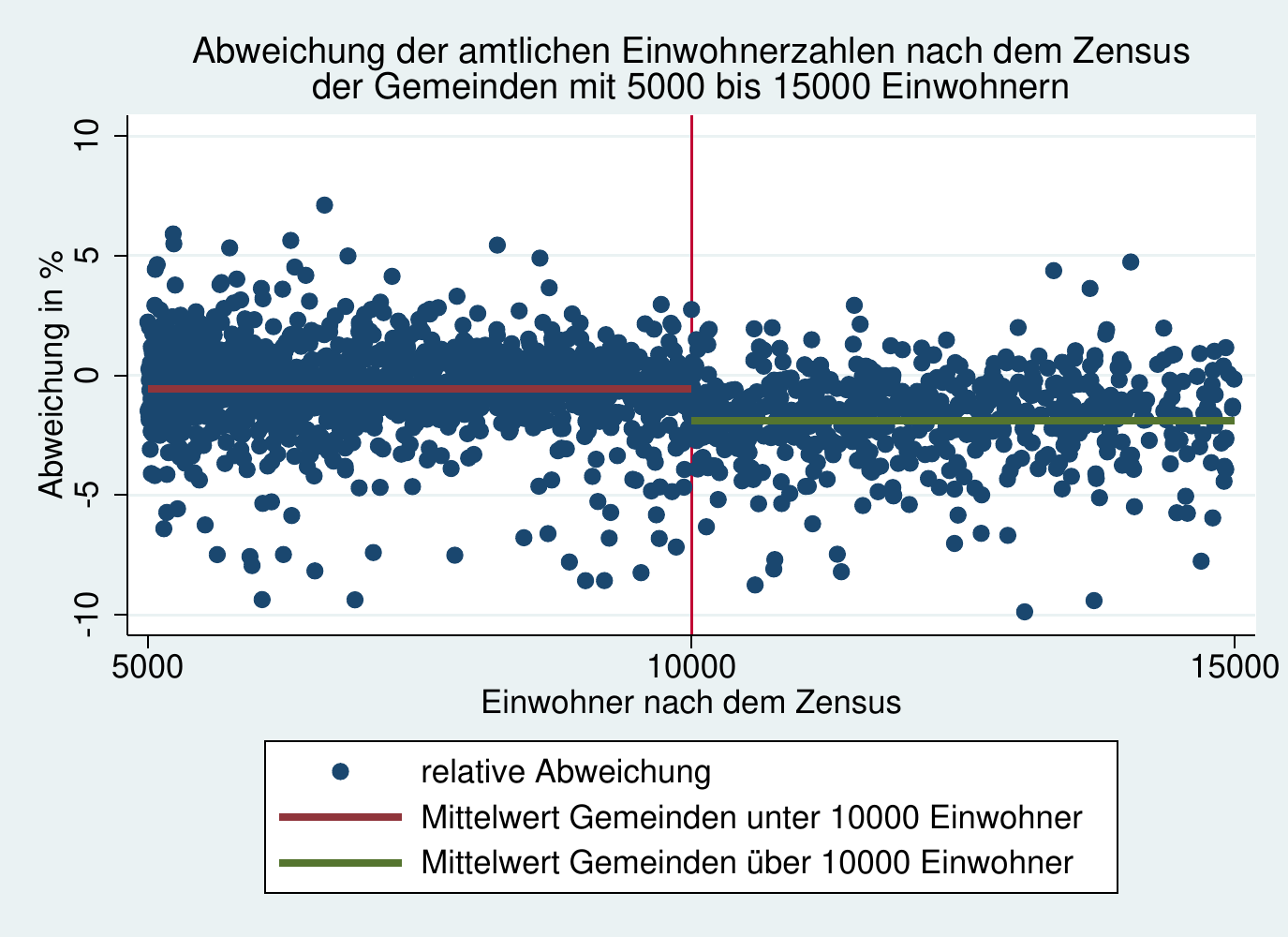}}
\subfigure[Intervall von 9.000 bis 11.000 Einwohner \label{abb4b}]{\includegraphics[width=7cm,height=7.5cm]{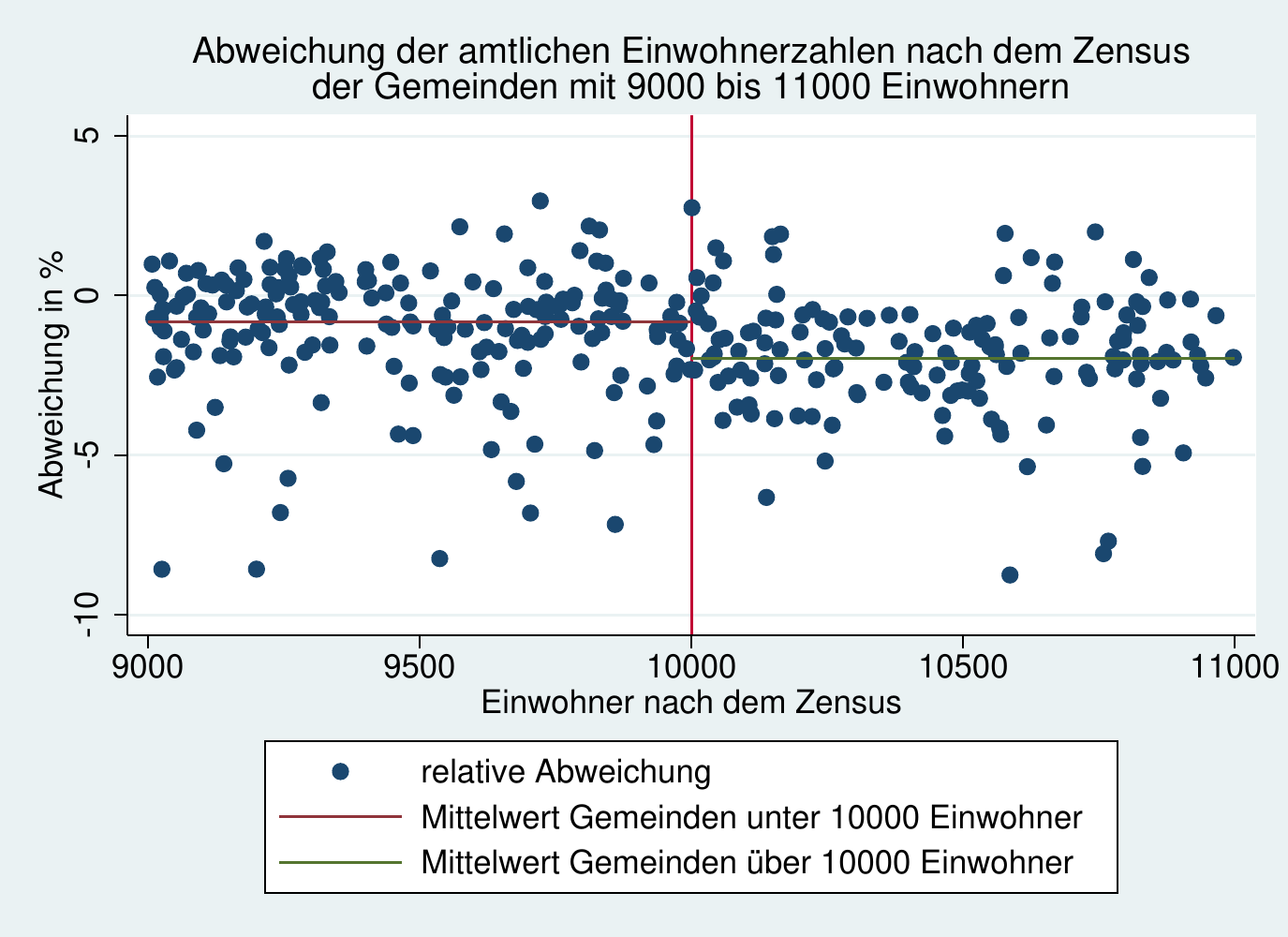}}
\end{center}
\end{figure}

Hinsichtlich der Mittelwertbetrachtung, der relativen Ver\"anderung der amtlichen Bev\"olkerungszahl nach alter Berechnung zur neuen Berechnung durch den Zensus mit Hilfe des Mann-Whitney-U-Tests, muss ausgeschlossen werden, dass weitere strukturelle Faktoren die Ergebnisse beeinflussen. M\"ogliche Einflussfaktoren k\"onnten zum Beispiel st\"adtisch und l\"andlich spezifische Eigenschaften sein, die sich bei einer betrachteten Einwohnerspanne von 1.000 bis \"uber eine Million Einwohner ergeben. Festzuhalten ist aber, dass sich diese Faktoren minimieren, wenn sich die Analyse auf sogenannte Nachbarn fokussiert. Den Begriff Nachbarn definieren wir \"uber die Einwohnerzahl. Hierf\"ur wurde das Untersuchungsintervall auf [5.000,15.000] und [9.000,11.000] Einwohner reduziert. In der Analyse befinden sich somit lediglich Gemeinden, die von der Struktur eher simultane Charakteristiken aufweisen. Die beiden Grafiken unter Abbildung~\ref{abb4} zeigen  Scatterplots der relativen Ver\"anderungen f\"ur alle Gemeinden aus den beiden Intervallen. Zus\"atzlich enthalten die Grafiken den Mittelwert f\"ur beide Gruppen.\\

Abbildung~\ref{abb4a} zeigt deutlich, dass f\"ur die von uns definierten Nachbargemeinden im Intervall [5.000,15.000] der Effekt bestehen bleibt, dass im Mittelwert Gemeinden \"uber 10.000 Einwohner schlechtere relative Ver\"anderungen der amtlichen Einwohnerzahl durch den Zensus 2011 aufweisen. Dieses Ergebnis bleibt bestehen, wenn wir das Intervall auf [9.000,11.000] verringern. Sowohl f\"ur das Intervall [5.000,15.000] als auch f\"ur das Intervall [9.000,11.000] haben wir mit Hilfe des Mann-Whitney-U-Tests auf statistische Signifikanz getestet. In beiden F\"allen zeigt sich, dass weiterhin Gemeinden \"uber 10.000 Einwohner signifikant schlechtere relative Ver\"anderungen in der Einwohnerzahl aufweisen (zweiseitiger Mann-Whitney-U-Test, $p<0.001$).\footnote{F\"ur das Land Sachsen-Anhalt ergibt sich ein \"ahnliches Ergebnis. Auch hier bleibt das Ergebnis, das Gemeinden ab 10.000 Einwohner schlechtere relative Ver\"anderungen aufweisen, signifikant, obwohl die Beobachtungszahl deutlich geringer ausf\"allt (zweiseitiger Mann-Whitney-U-Test, $p<0.05$).} \footnote{Um unsere Argumentationslinie zu untermauern haben wir zus\"atzlich eine Analyse der Gemeinden unter 10.000 Einwohner durchgef\"uhrt, in der wir die Gruppe in zwei Untergruppen unterteilt haben. Als Trenngrenze wurde der Wert 5.000 Einwohner gew\"ahlt. Untersucht man dann mit Hilfe des Mann-Whitney-U-Test diese zwei Gruppen, so l\"asst sich kein signifikanter Unterschied in der relativen Ver\"anderung der Einwohnerzahl finden. Dieses Ergebnis bleibt auch bestehen, wenn man die Trenngrenze auf andere Werte innerhalb des Intervalls von 1.000 bis unter 10.000 Einwohner setzt.} Somit wird deutlich, dass weitere strukturelle Faktoren keinen elementaren Effekt haben k\"onnen, um die von uns im Kapitel~\ref{K4} aufgezeigten Ergebnisse zu beeinflussen. Vielmehr werden unsere Erkenntnisse durch die Nachbar-Analyse noch best\"atigt. \\

Im Kapitel~\ref{K4} wurden bereits Analysen mit Hilfe einer multiplen Regression durchgef\"uhrt. Um nun noch weitere Aussagen treffen zu k\"onnen, wird das Regressionsmodell um zus\"atzliche exogene Variablen erweitert. Dabei handelt es sich um den Ausl\"anderanteil, den Studentenanteil der Studenten an der Hauptwohnsitzbev\"olkerung sowie einer Gendervariable, die das Verh\"altnis zwischen m\"annlicher und weiblicher Bev\"olkerung widerspiegelt. Tabelle~\ref{Tab2} zeigt die Ergebnisse der Regressionsanalysen.\\ 

\begin{table}[h]
\caption{Ergebnisse der Regressionen Teil 2}\label{Tab2}
\begin{center}
\scalebox{0.80}{%
\begin{tabular}{lcc} \hline
 & (1) & (2)  \\
VARIABLEN & Relative Ver\"anderung  & Relative Ver\"anderung    \\ 
& Sachsen-Anhalt & Deutschland  \\ \hline
Methode & -1.525*** & -1.476***  \\
 & (0.244) & (0.0796)  \\
Einwohner & 2.49e-06 & -6.40e-07 \\
 & (4.48e-06) & (5.09e-07)  \\
Ausl\"anderquote & -13.74** & -0.497   \\
 & (6.490) & (0.796)   \\
Gender & 6.916 & 10.55***  \\
 & (9.333) & (2.491)  \\
Studentenanteil & -4.639 & -3.708**  \\
 & (9.896) & (1.464)  \\
Constant & -3.812 & -5.858***  \\
 & (4.667) & (1.250)  \\
Beobachtungen & 201 & 7148  \\
$R^2$ & 0.283 & 0.060  \\
 Adj. $R^2$ & 0.264 & 0.0596  \\ \hline
\multicolumn{3}{c}{Standardfehler in Klammern} \\
\multicolumn{3}{c}{ *** p$<$0,01, ** p$<$0,05, * p$<$0,1} \\
\end{tabular}}

\end{center}
\end{table}

Wie aus Tabelle~\ref{Tab2} ersichtlich wird, bleibt die exogene Variable Methode in beiden Regressionen weiterhin signifikant. Dies ist gleichbedeutend damit, dass die Hypothese, dass Gemeinden \"uber 10.000 Einwohner aufgrund der unterschiedlich verwendeten Methoden signifikant schlechter gestellt werden, robust auch unter der Hinzunahme von weiteren exogenen Variablen bleibt. Auch \"andern sich das Vorzeichen sowie die Signifikanzbetrachtung nicht f\"ur die Variable Einwohner. Hinsichtlich der neuen Variablen l\"asst sich zeigen, dass bei der Betrachtung der Daten f\"ur den Bund, die Anzahl der Studenten zur Hauptwohnsitzbev\"olkerung einen signifikant negativen Effekt aufweist. Ein gleicher Effekt ergibt sich f\"ur die Ausl\"anderquote, wobei hier das Ergebnis nicht statistisch signifikant ist. Die Gendervariable weist dagegen einen positiven Regressionskoeffizienten auf, der zudem statistisch signifikant ist.\footnote{F\"ur Sachsen-Anhalt l\"asst sich lediglich beim Ausl\"anderanteil ein signifikanter Koeffizient finden, die Koeffizienten f\"ur die Gendervariable sowie die Studentenquote sind dagegen nicht signifikant. Die Regressionen f\"ur die anderen Bundesl\"ander lasse sich im Anhang finden.} Das bedeutet, dass Gemeinden die einen h\"oheren Anteil an Frauen innerhalb ihrer Bev\"olkerung aufweisen, signifikant bessere Zensusergebnisse erhalten haben.\footnote{Dieser Effekt bleibt bestehen, wenn man sich nur die Gemeinden mit Stichprobe anschaut. Im Umkehrschluss bedeutet dies, dass innerhalb der Stichprobe h\"aufiger die Frauen auch angetroffen wurden.}  \\

\begin{figure}[h]
\caption{Regressionsgerade konsolidiertes Melderegister zu bisheriger amtlicher Einwohnerzahl}\label{abb6}
\begin{center}
 \includegraphics[width=12cm,height=9cm]{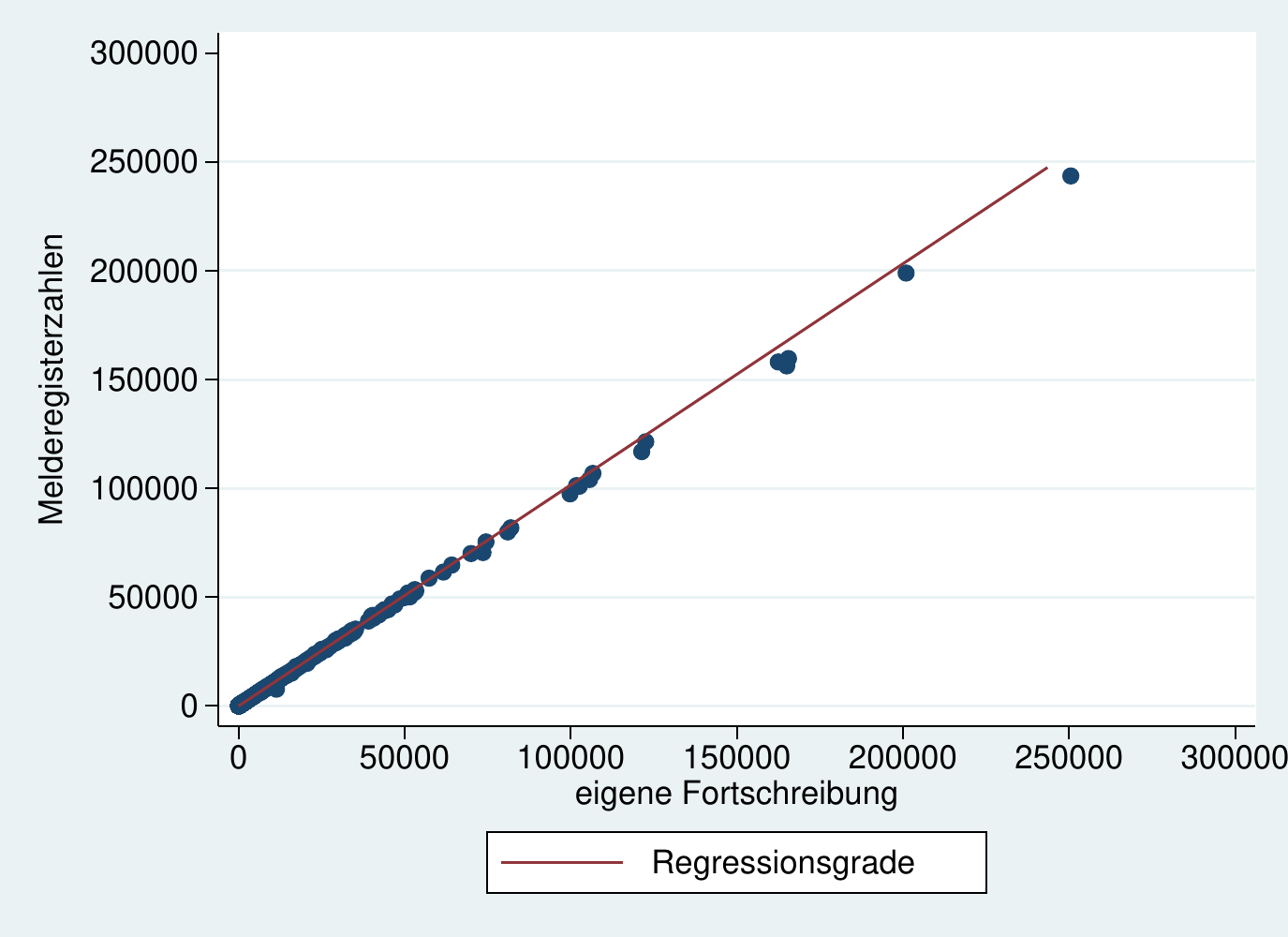}
 \end{center}
\end{figure}

Ein Nebenprodukt des Zensus 2011 war es die Qualit\"at der Melderegister im Vergleich zu den amtlichen Einwohnerzahlen zu analysieren. Zur Berechnung der neuen amtlichen Bev\"olkerungszahlen wurden in mehrstufigen Verfahren die konsolidierten Melderegisterbest\"ande ermittelt. Insbesondere der konsolidierte Melderegisterbestand vom 09.05.2011 hat hierbei eine hohe Relevanz. Im folgenden sollen die bisherigen Ergebnisse unter Ber\"ucksichtigung der konsolidierten Melderegisterbest\"ande betrachtet werden. Leider liegen die konsolidierten Melderegisterbest\"ande nur f\"ur die Bundesl\"ander Niedersachsen, Rheinland-Pfalz sowie Baden-W\"urttemberg vor. Um zeigen zu k\"onnen, dass unsere Untersuchungen mit den bisherigen amtlichen Einwohnerzahlen eine hohe Validit\"at aufweisen, haben wir diese Zahlen f\"ur die 3 Bundesl\"ander zu den dazugeh\"origen konsolidierten Melderegisterzahlen analysiert. Hierf\"ur haben wir eine einfache lineare Regression genutzt. Die Abbildung~\ref{abb6} zeigt hierf\"ur die Regressionsgerade. \\

Wie man aus Abbildung~\ref{abb6} leicht erkennen kann, liegen die Datenpunkte und die daraus resultierende Regressionsgerade nahezu auf der 45-Grad-Linie durch das Koordinatensystem. Das bedeutet, dass die konsolidierten Melderegisterbest\"ande sehr nah an dem bisherigen amtlichen Bev\"olkerungszahlen liegen. Dies wird durch die Ergebnisse der Regression unterst\"utzt. Der Regressionskoeffizient liegt mit 0,9836 nah an eins. Das bedeutet, dass die konsolidierten Melderegisterbest\"ande im Vergleich zu den bisherigen amtlichen Einwohnerzahlen bei 98\% liegen. Diese Regressionsanalyse weist ein hohes Bestimmtheitsmaß von 0,99 auf. Entsprechend kann man zu dem Ergebnis kommen, dass die amtliche Bev\"olkerungszahl ein sehr guter Proxy f\"ur die konsolidierten Melderegisterzahlen ist und bei Nicht-Vorliegen dieser Daten ohne weiteres auf die bisherigen amtlichen Einwohnerzahlen f\"ur Zensusanalysen zur\"uckgegriffen werden kann. \\

Die konsolidierten Melderegisterdaten haben wir wie die bisherigen amtlichen Fortschreibungszahlen zus\"atzlich allen statistischen Untersuchungen unterzogen. Konkret bedeutet dies Untersuchungen nach dem Mann-Whitney-U-Test sowie die zwei durchgef\"uhrten linearen Regressionen. F\"ur die nicht-parametrischen Untersuchungen zeigt sich, dass in allen drei Bundesl\"andern die relativen Ver\"anderungen signifikant schlechter f\"ur die Gruppe der Gemeinden ab 10.000 Einwohner ausfallen (zweiseitiger Mann-Whitney-U-Test, $p<0.001$). Die einfache lineare Regression wie in Tabelle~\ref{Tab1} best\"atigt die Erkenntnisse f\"ur die drei Bundesl\"ander.\footnote{Die ausf\"uhrlichen Regressionsergebnisse f\"ur alle drei Bundesl\"ander sind im Anhang zu finden.} Die Regressionskoeffizienten ver\"andern sich nur marginal, das Signifikanzniveau der Dummyvariable Methode bleibt dagegen weiterhin hoch signifikant. Auch die Regressionsanalyse wie in Tabelle~\ref{Tab2} auf Grundlage der konsolidierten Melderegisterdaten zeigt die Problematik der Methodengrenze bei 10.000 Einwohner weiter auf. Die Signifikanz der Regressionskoeffizienten bleibt weiterhin hoch.\\

Die Ergebnisse zeigen also deutlich, dass durch die Nutzung der Melderegisterdaten anstelle der Daten der Bev\"olkerungsfortschreibung keine Ver\"anderungen in den wesentlichen Ergebnissen zutage treten, d.h. der Strukturbruch durch den Methodenwechsel bei Gemeinden ab 10.000 Einwohnern liegt auch in diesen Ergebnissen klar vor.\footnote{Die Argumentation, dass aufgrund der Verwendung der Daten der Bev\"olkerungsfortschreibung ein Strukturbruch auftreten k\"onnte, erscheint auch wenig plausibel, da dieses voraussetzen w\"urde, dass in der Vergangenheit bei der Berechnung der Daten der Bev\"olkerungsfortschreibung genau bei 10.000 Einwohnern ein Strukturbruch erzeugender Fehler aufgetreten sein m\"usste.}

\section{Conclusion} \label{K5}  
Auf Grundlage der durch das Statistische Bundesamt ver\"offentlichten neuen amtlichen Einwohnerzahlen des Zensus 2011 analysiert diese Arbeit die Ergebnisse im Vergleich zu den bisherigen amtlichen Einwohnerzahlen zum Stichtag 31.12.2011. Hauptaugenmerk liegt dabei auf der Tatsache, dass f\"ur Gemeinden unter 10.000 Einwohner und ab 10.000 Einwohner unterschiedliche Methoden zur Ermittlung der amtlichen Einwohnerzahl verwendet wurden. Die Untersuchung der Verteilungen der relativen Ver\"anderung der Einwohnerzahl zwischen der neuen und alten Berechnung zeigen deutlich auf, dass Gemeinden unter 10.000 Einwohner vom Zensus 2011 profitieren. \\

Um \"uberpr\"ufen zu k\"onnen, ob die unterschiedlich verwendeten Methoden oder eventuell, die m\"ogliche schlechtere Qualit\"at der Melderegister in gr\"oßeren St\"adten die Ursache hierf\"ur sind, wurden die Zensusergebnisse zus\"atzlich mit Hilfe von Regressionsanalysen untersucht. Dabei zeigt sich, dass ein zus\"atzlicher gemeldeter Einwohner je Gemeinde keinen Effekt auf die Ver\"anderung der amtlichen Einwohnerzahl zwischen alter und neuer Berechnung hat. Das bedeutet somit auch, dass die Qualit\"at der Melderegister sich nicht hinsichtlich der Gemeindegr\"oße signifikant unterscheidet bzw. im Umkehrschluss, dass auch gr\"oßere St\"adte sehr wohl in der Lage sind gut gef\"uhrte Register aufzuweisen.\\

Die Ergebnisse der Regressionsanalyse zeigen jedoch deutlich auf, dass die unterschiedlichen Methoden, die zur Berechnung der amtlichen Bev\"olkerung Verwendung finden, sehr wohl einen signifikanten Effekt aufweisen. So werden Gemeinden ab 10.000 Einwohner, in denen Haushaltsbefragungen auf Stichprobenbasis als Grundlage zur Berechnung der amtlichen Einwohnerzahl verwendet wurden, benachteiligt. Mit dem Beschluss der Bundesregierung am Zensus 2011 teilzunehmen und gleichzeitig unterschiedliche Methoden zur Berechnung der amtlichen Einwohnerzahl zu verwenden, muss die Nebenbedingung existiert haben, dass es durch die Nutzung unterschiedlicher Methoden, zu keinen signifikanten Benachteiligungen von bestimmten Gemeindegruppen kommen darf. \\

Bemerkenswert ist, dass unsere Erkenntnisse auch unter tiefer gehenden Analysen bestehen bleiben. Sowohl f\"ur die Nachbarschaftsanalysen als auch f\"ur die Regressionen mit zus\"atzlichen exogenen Variablen zeigt sich, dass die Methode einen signifikanten und strukturellen Unterschied zur Folge hat. Die in dieser Arbeit aufgef\"uhrten Ergebnisse zeigen allerdings, dass gegen diese Nebenbedingung verstoßen wird.\\

Auch bleiben die Ergebnisse robust gegen\"uber der Verwendung von konsolidierten Melderegisterzahlen zum 9. Mai 2011. Zwar liegen diese Daten nur f\"ur die drei Bundesl\"ander, Niedersachsen, Rheinland-Pfalz und Baden-W\"urttemberg vor, aber auch f\"ur diese kleinere Stichprobe k\"onnen wir unsere Ergebnisse reproduzieren. Somit zeigt diese Untersuchung deutlich, dass der Zensus 2011 durch die unterschiedliche Methodenverwendung zur Berechnung der neuen amtlichen Einwohnerzahl Gewinner und Verlierer hervorgebracht hat.


\pagebreak

\begin{appendix}

\section{Anhang}

\begin{figure}[h]
\caption{Scatterplot bisherige amtliche Einwohnerzahl zu den relativen Ver\"anderungen Baden-W\"urttemberg}\label{Anhang1}
\begin{center}
 \includegraphics[width=9.5cm,height=8cm]{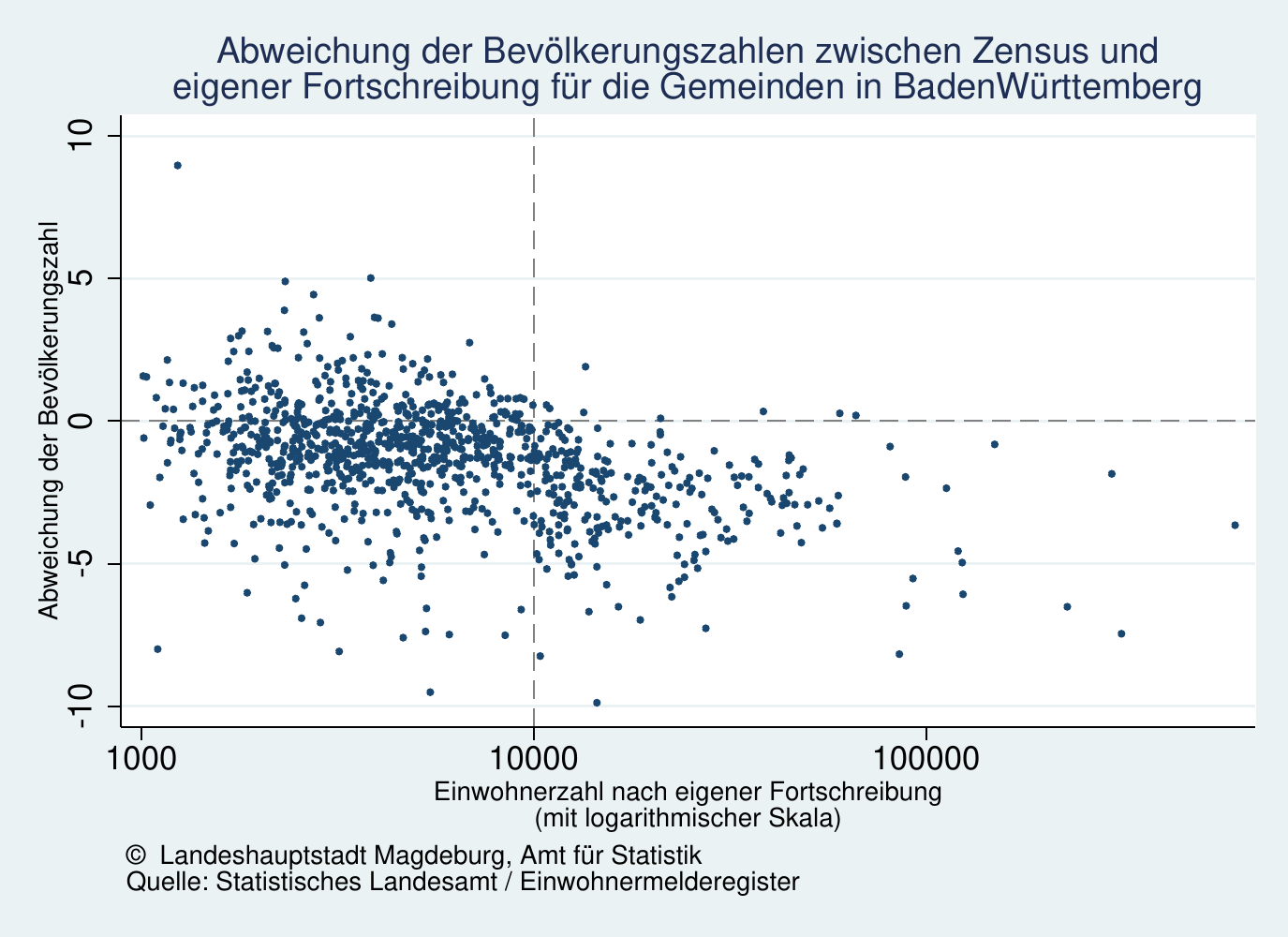}
 \end{center}
\caption{Scatterplot bisherige amtliche Einwohnerzahl zu den relativen Ver\"anderungen Bayern}\label{Anhang2}
\begin{center}
 \includegraphics[width=9.5cm,height=8cm]{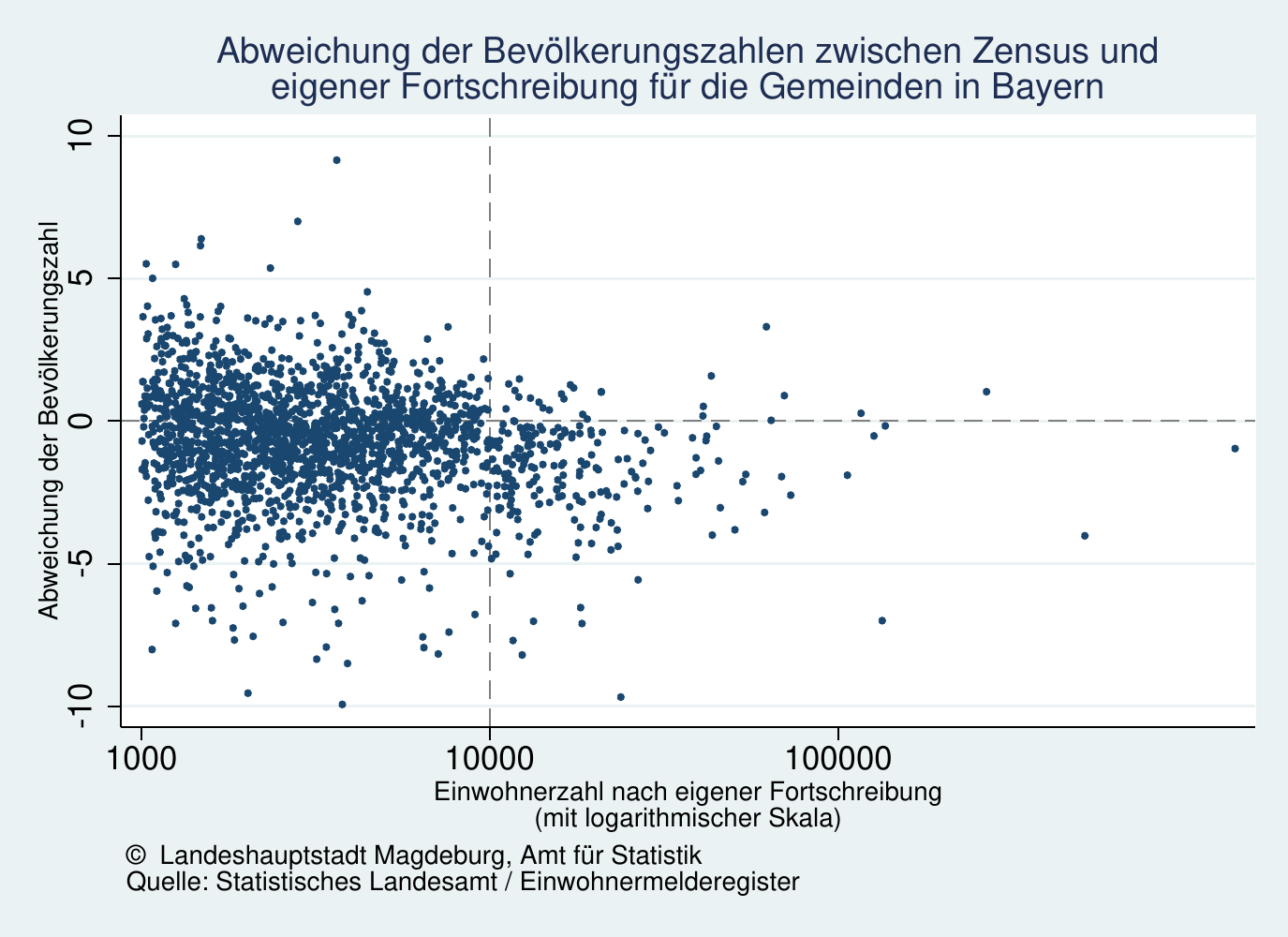}
 \end{center}
\end{figure}

\begin{figure}[h]
\caption{Scatterplot bisherige amtliche Einwohnerzahl zu den relativen Ver\"anderungen Brandenburg}\label{Anhang4}
\begin{center}
 \includegraphics[width=9.5cm,height=8cm]{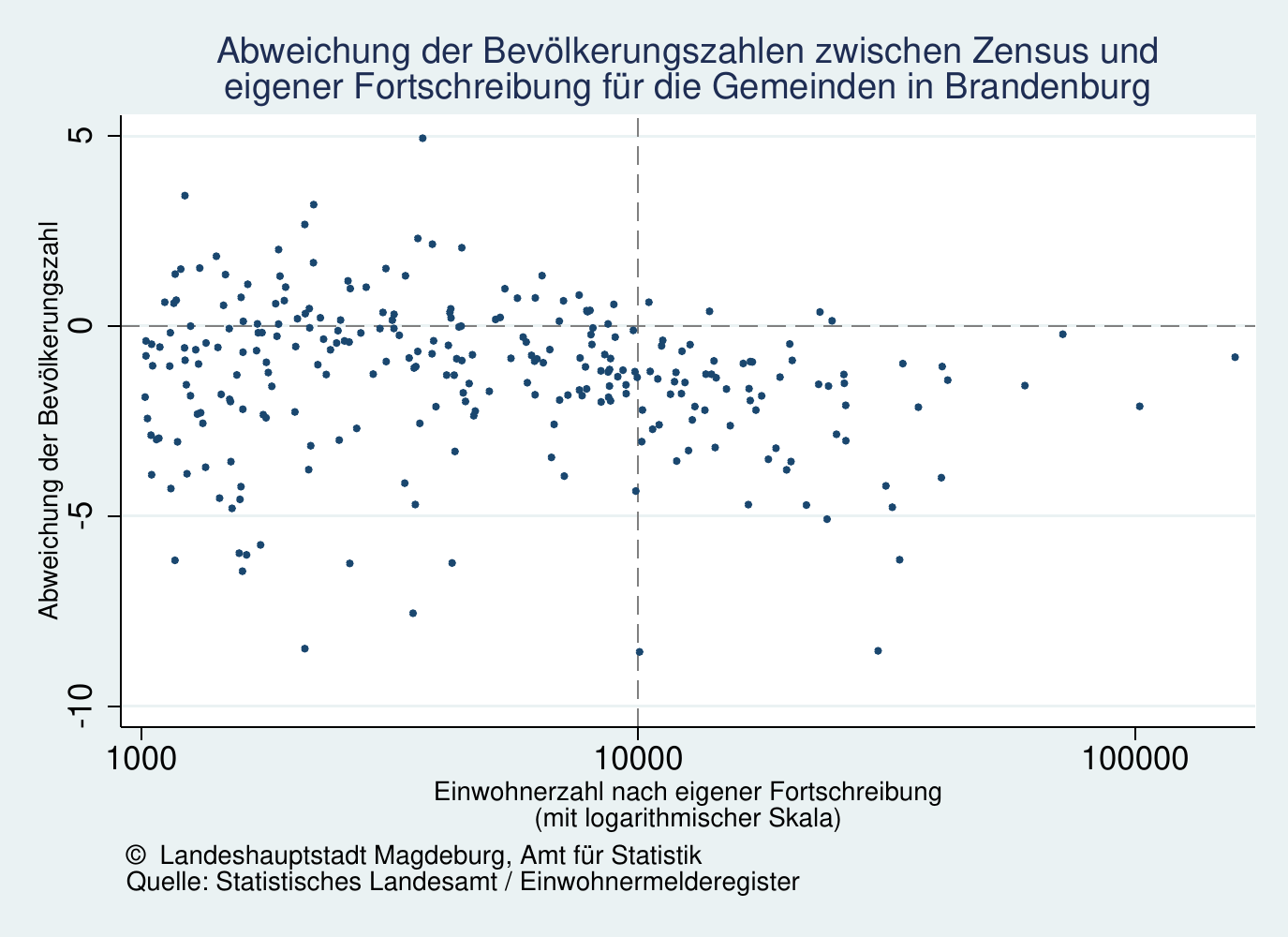}
 \end{center}
\caption{Scatterplot bisherige amtliche Einwohnerzahl zu den relativen Ver\"anderungen Hessen}\label{Anhang7}
\begin{center}
 \includegraphics[width=9.5cm,height=8cm]{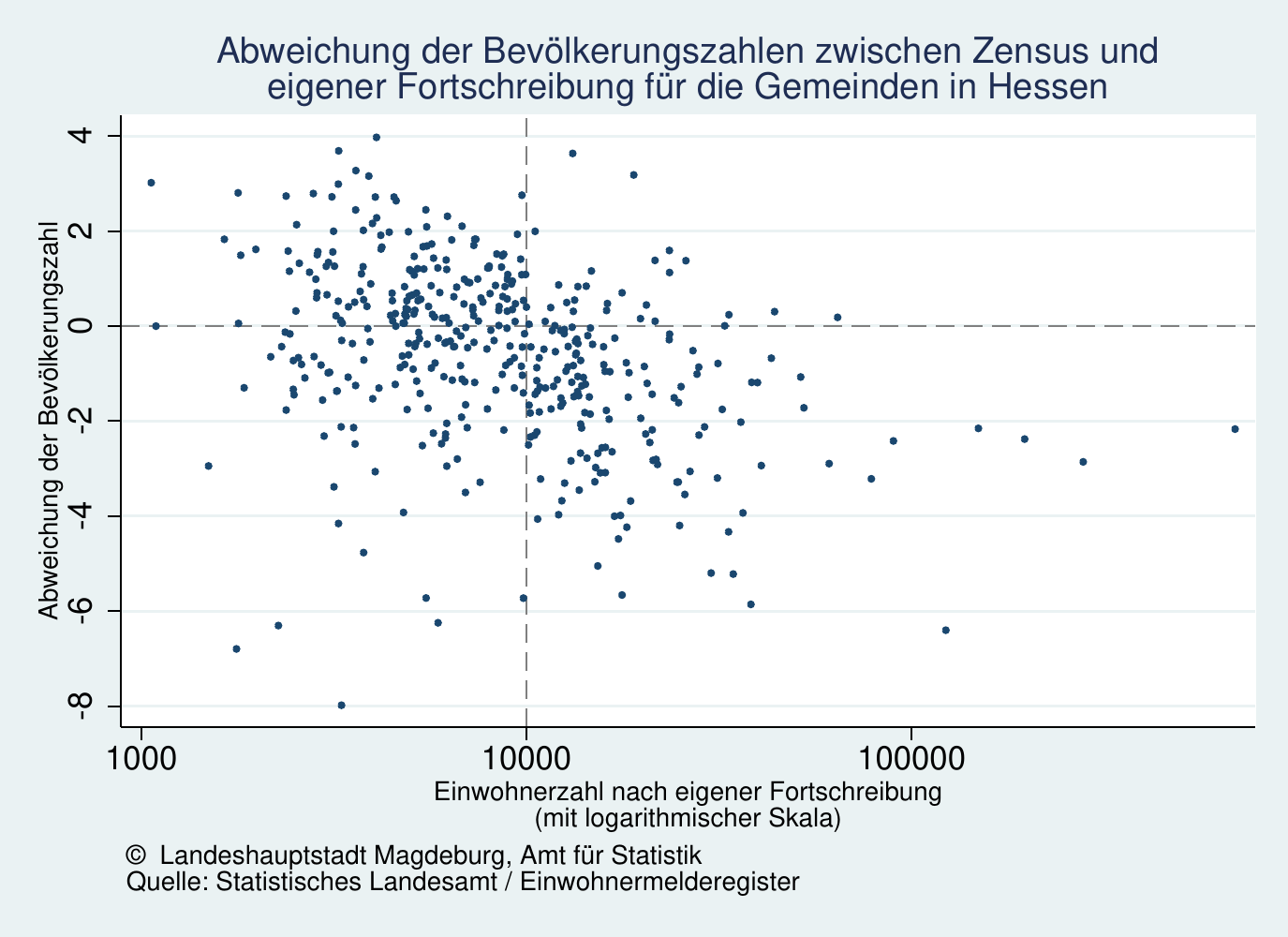}
 \end{center}
\end{figure}

\begin{figure}[h]
\caption{Scatterplot bisherige amtliche Einwohnerzahl zu den relativen Ver\"anderungen Mecklenburg-Vorpommern}\label{Anhang8}
\begin{center}
 \includegraphics[width=9.5cm,height=8cm]{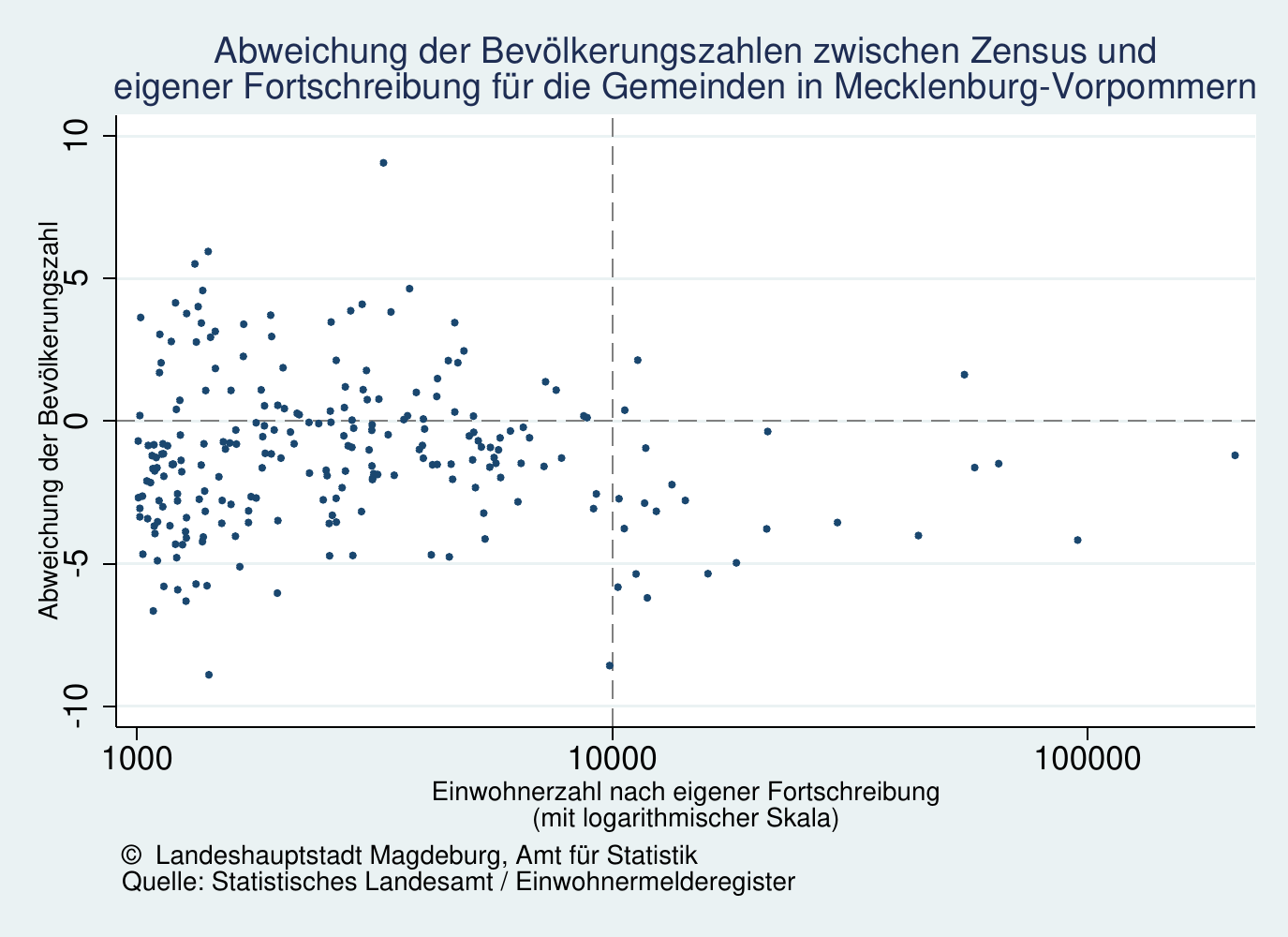}
 \end{center}
\caption{Scatterplot bisherige amtliche Einwohnerzahl zu den relativen Ver\"anderungen Niedersachsen}\label{Anhang9}
\begin{center}
 \includegraphics[width=9.5cm,height=8cm]{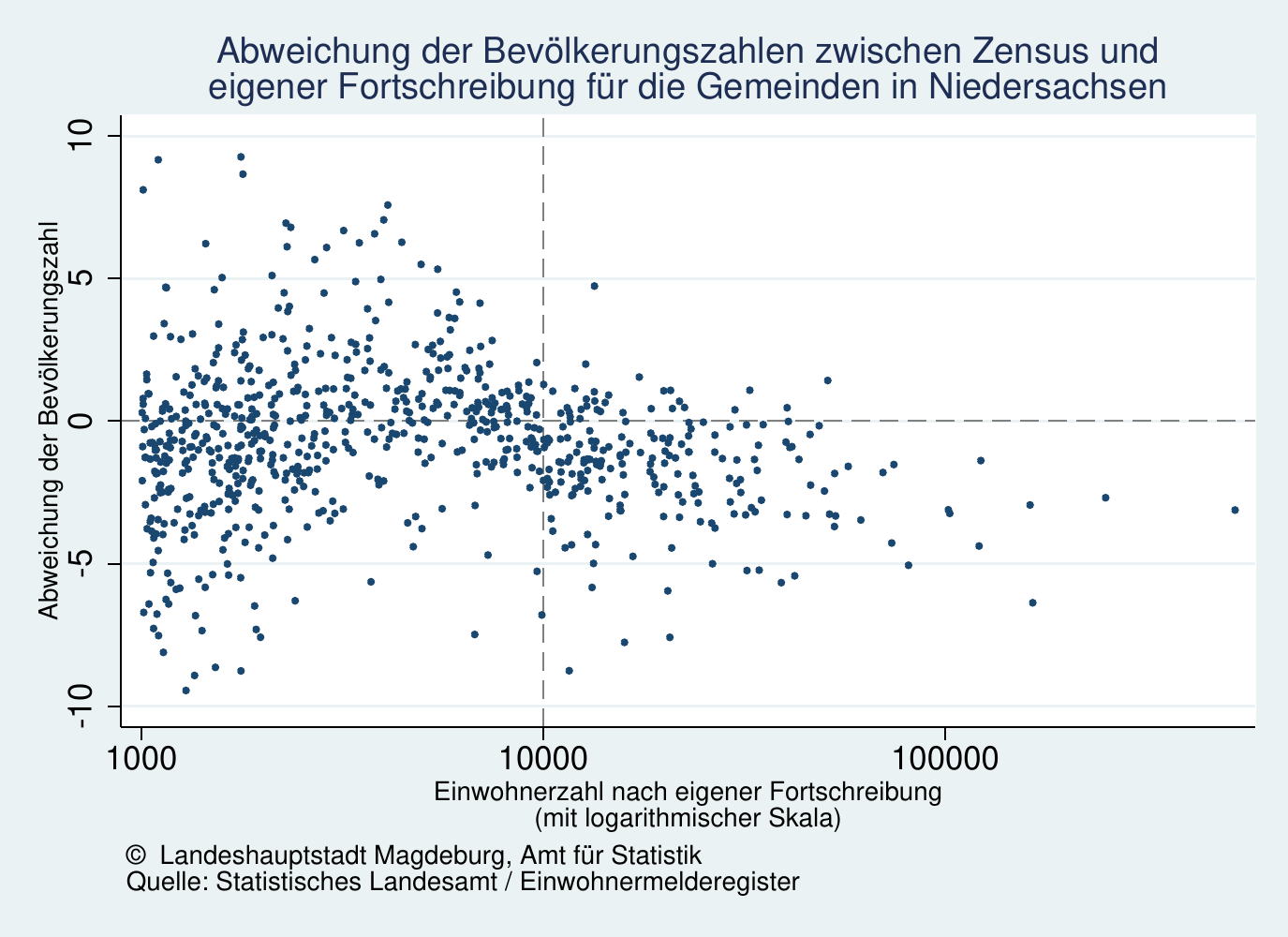}
 \end{center}
\end{figure}

\begin{figure}[h]
\caption{Scatterplot bisherige amtliche Einwohnerzahl zu den relativen Ver\"anderungen Nordrhein-Westfalen}\label{Anhang10}
\begin{center}
 \includegraphics[width=9.5cm,height=8cm]{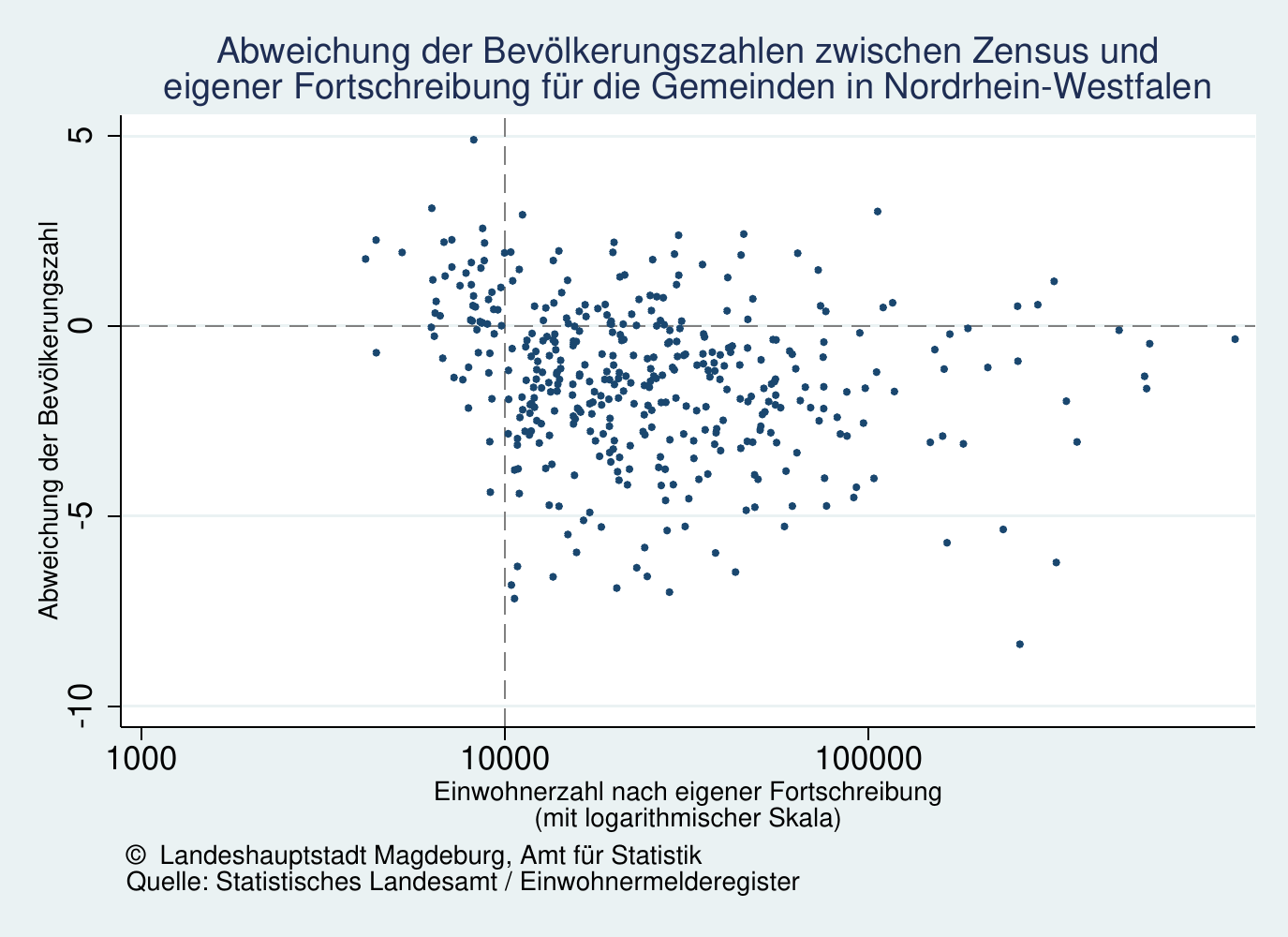}
 \end{center}
\caption{Scatterplot bisherige amtliche Einwohnerzahl zu den relativen Ver\"anderungen Rheinland-Pfalz}\label{Anhang11}
\begin{center}
 \includegraphics[width=9.5cm,height=8cm]{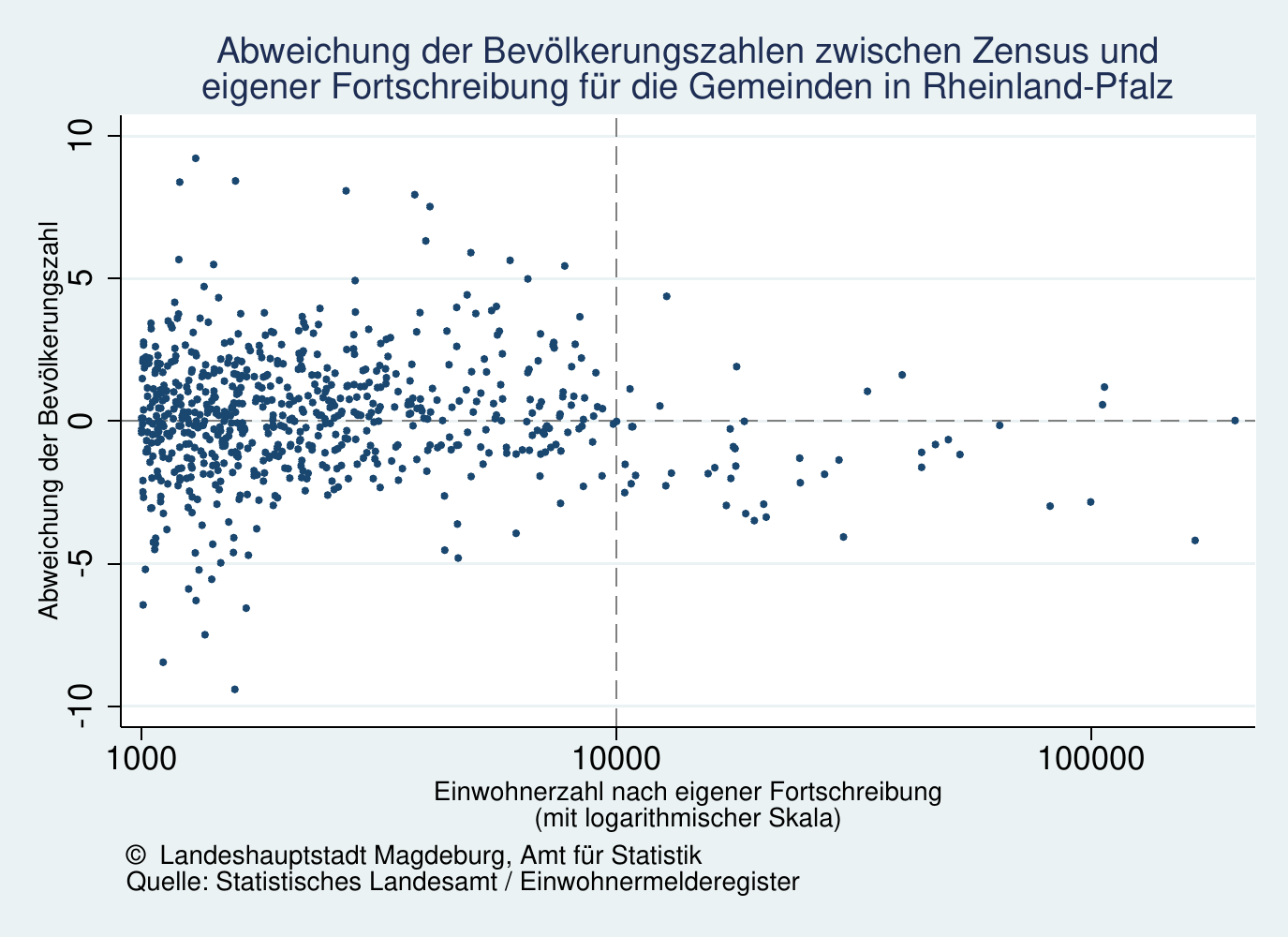}
 \end{center}
\end{figure}

\begin{figure}[h]
\caption{Scatterplot bisherige amtliche Einwohnerzahl zu den relativen Ver\"anderungen Saarland}\label{Anhang12}
\begin{center}
 \includegraphics[width=9.5cm,height=8cm]{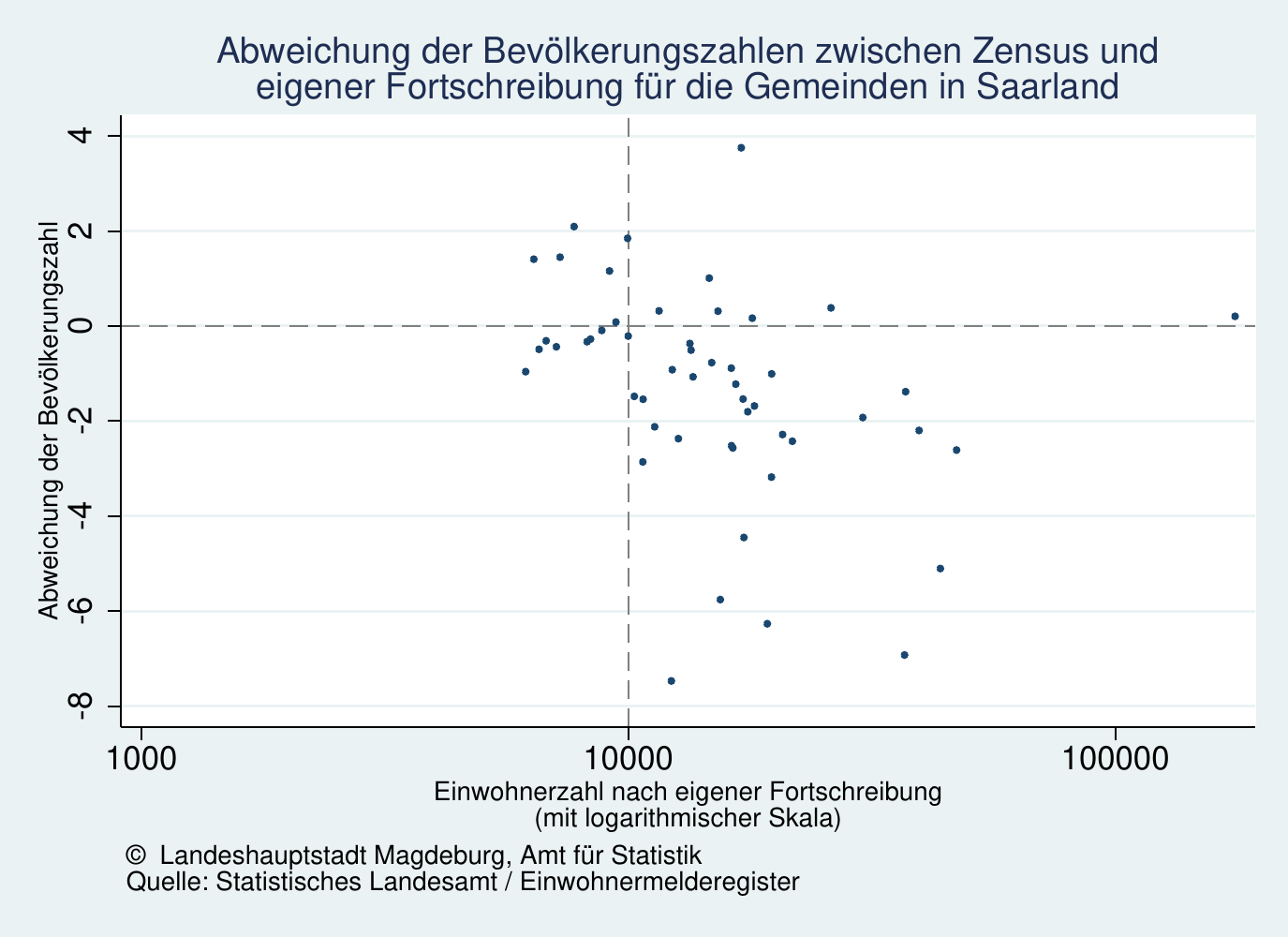}
 \end{center}
\caption{Scatterplot bisherige amtliche Einwohnerzahl zu den relativen Ver\"anderungen Sachsen}\label{Anhang13}
\begin{center}
 \includegraphics[width=9.5cm,height=8cm]{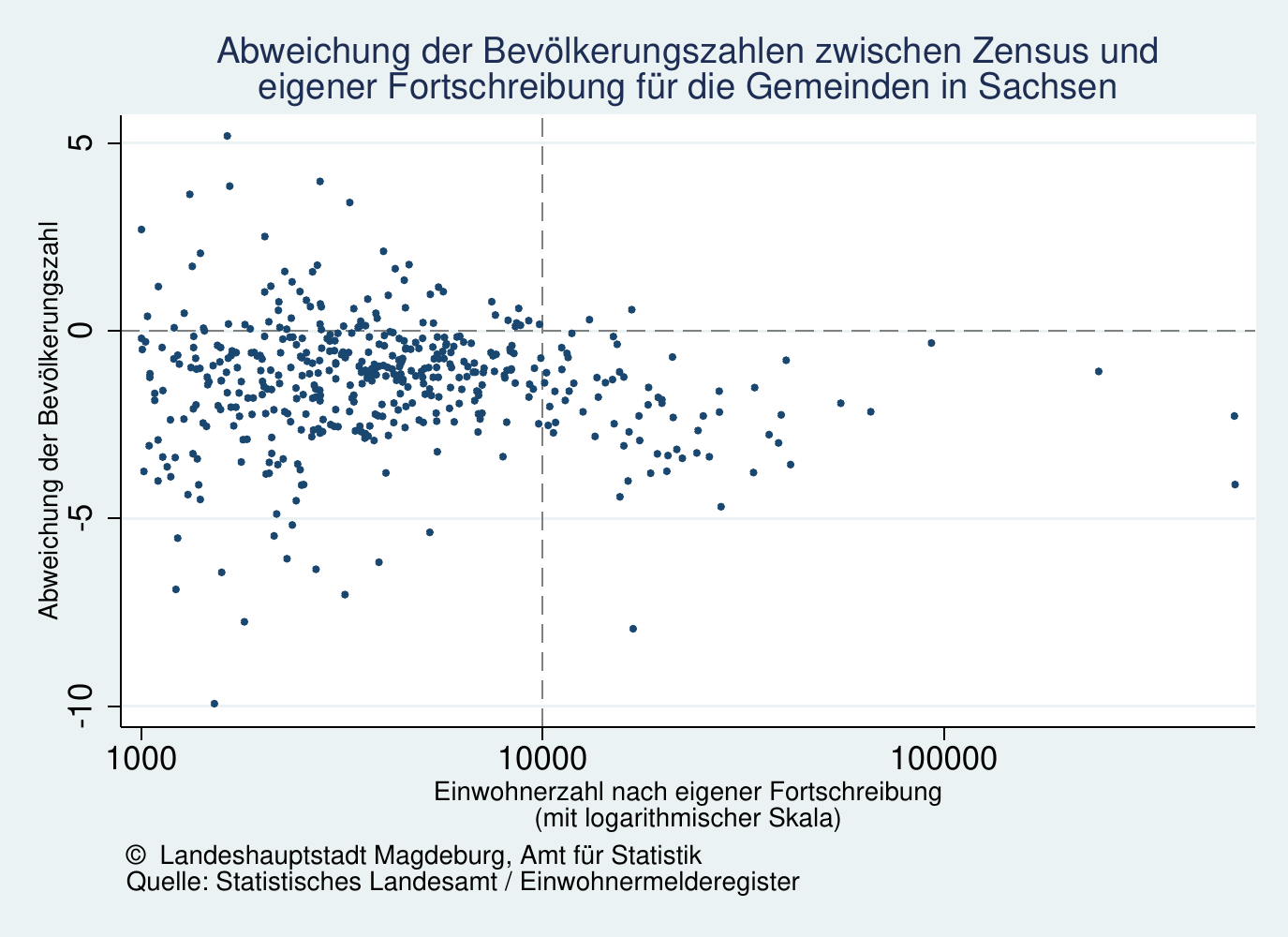}
 \end{center}
\end{figure}

\begin{figure}[h]
\caption{Scatterplot bisherige amtliche Einwohnerzahl zu den relativen Ver\"anderungen Schleswig-Holstein}\label{Anhang14}
\begin{center}
 \includegraphics[width=9.5cm,height=8cm]{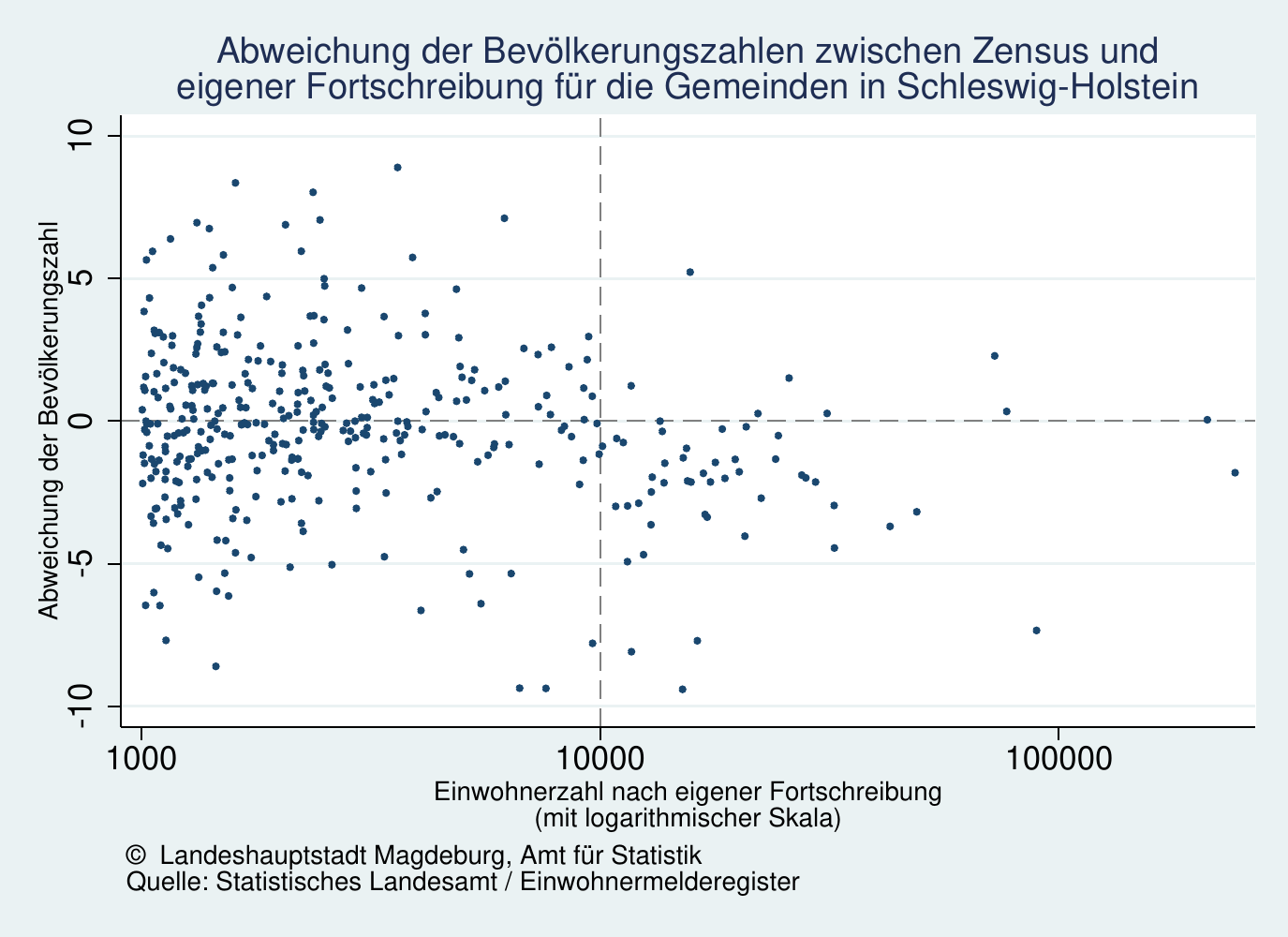}
 \end{center}
\caption{Scatterplot bisherige amtliche Einwohnerzahl zu den relativen Ver\"anderungen Th\"uringen}\label{Anhang15}
\begin{center}
 \includegraphics[width=9.5cm,height=8cm]{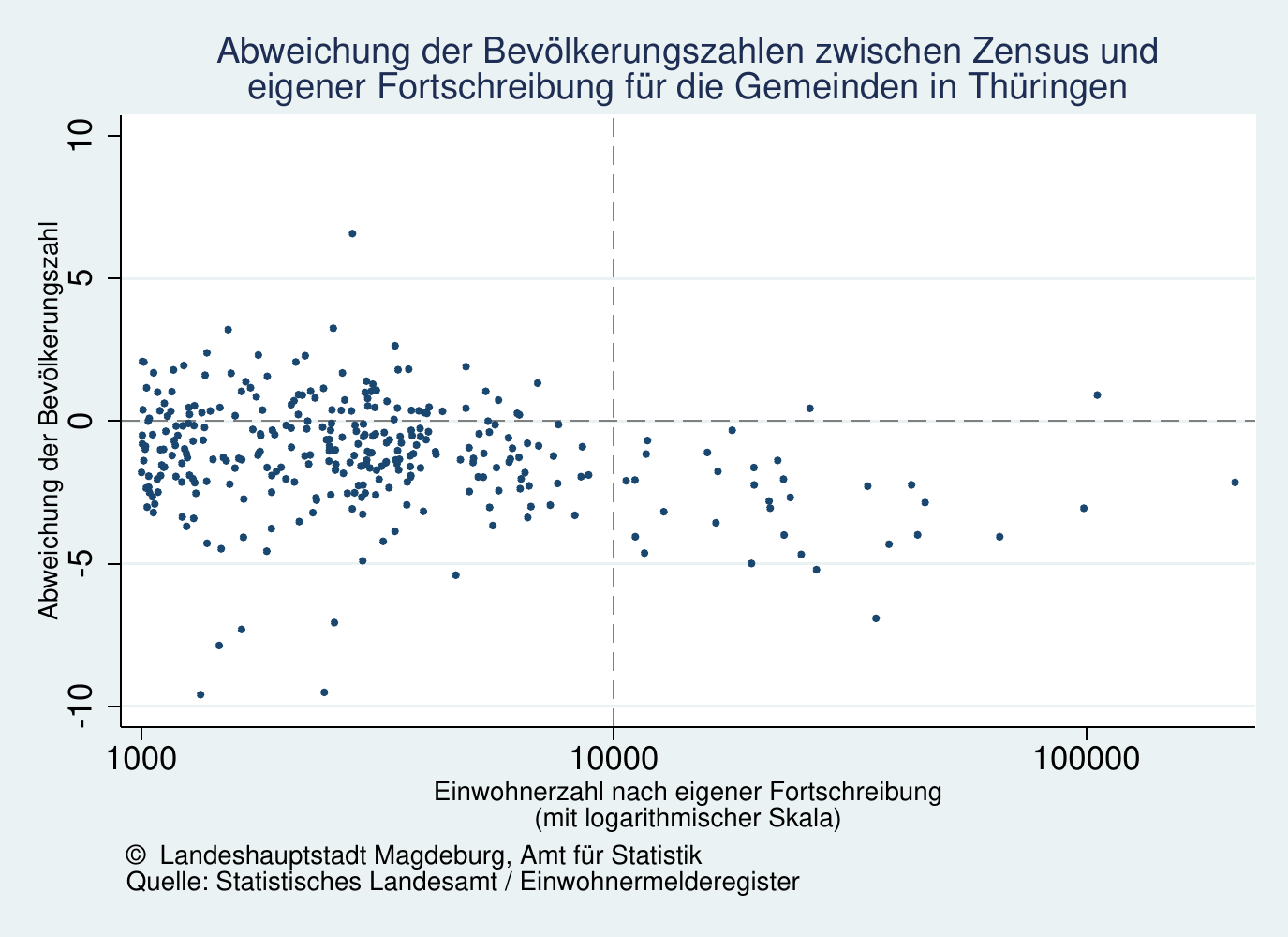}
 \end{center}
\end{figure}

\begin{sidewaystable}

\caption{Ergebnisse der Regressionen f\"ur die Bundesl\"ander Teil 1a}\label{Tab3a}
\begin{center}
\scalebox{0.70}{%
\begin{tabular}{lcccccccccccc} \hline
 & (1) & (2) & (3) & (4) & (5) & (6)  \\
VARIABLEN & Relative Ver\"anderung & Relative Ver\"anderung & Relative Ver\"anderung & Relative Ver\"anderung & Relative Ver\"anderung & Relative Ver\"anderung \\ 
 & Baden-W\"urttemberg & Bayern  & Brandenburg  & Hessen  & Mecklenburg-Vorpommern  & Niedersachsen    \\ \hline
Methode & -1.700*** & -1.052*** & -1.178*** & -1.477*** & -1.946** & -1.442***  \\
 & (0.159) & (0.158) & (0.355) & (0.186) & (0.752) & (0.277)  \\
Einwohner & -5.59e-06** & 2.65e-07 & 5.41e-06 & -4.30e-06* & 1.11e-05 & -4.13e-06  \\
 & (2.47e-06) & (1.43e-06) & (1.08e-05) & (2.33e-06) & (1.38e-05) & (4.76e-06) ) \\
Constant & -0.918*** & -0.694*** & -1.114*** & 0.0384 & -1.128*** & -0.374***  \\
 & (0.0724) & (0.0519) & (0.148) & (0.112) & (0.198) & (0.130)  \\

Beobachtungen & 1027 & 1905 & 273 & 424 & 236 & 765  \\
$R^2$ & 0.139 & 0.024 & 0.051 & 0.162 & 0.030 & 0.049  \\
 Adj. $R^2$ & 0.137 & 0.0230 & 0.0441 & 0.158 & 0.0221 & 0.0468  \\ \hline
\multicolumn{7}{c}{Standardfehler in Klammern} \\
\multicolumn{7}{c}{ *** p$<$0,01, ** p$<$0,05, * p$<$0,1} \\
\end{tabular}}
\end{center}
\end{sidewaystable}

\begin{sidewaystable}

\caption{Ergebnisse der Regressionen f\"ur die Bundesl\"ander Teil 1b}\label{Tab3b}
\begin{center}
\scalebox{0.70}{%
\begin{tabular}{lcccccccccccc} \hline
 & (7) & (8) & (9) & (10) & (11) & (12) \\
VARIABLEN & Relative Ver\"anderung & Relative Ver\"anderung & Relative Ver\"anderung & Relative Ver\"anderung & Relative Ver\"anderung & Relative Ver\"anderung  \\ 
 &  Nordrhein-Westfalen  & Rheinland-Pfalz  & Saarland  & Sachsen  & Schleswig-Holstein  & Th\"uringen    \\ \hline
Methode &  -1.897*** & -1.501*** & -2.356*** & -0.862*** & -2.788*** & -1.811*** \\
 &  (0.319) & (0.477) & (0.657) & (0.230) & (0.758) & (0.473) \\
Einwohner &  2.03e-07 & 1.99e-06 & 5.22e-06 & -1.25e-06 & 8.48e-06 & 2.74e-06 \\
 &  (1.25e-06) & (8.73e-06) & (1.22e-05) & (2.19e-06) & (1.43e-05) & (9.28e-06) \\
Constant &  0.143 & 0.282*** & 0.311 & -1.237*** & -0.0811 & -1.002*** \\
 &  (0.292) & (0.0973) & (0.546) & (0.0836) & (0.244) & (0.122) \\

Beobachtungen &  396 & 693 & 52 & 457 & 391 & 324 \\
$R^2$ &  0.084 & 0.021 & 0.213 & 0.038 & 0.039 & 0.063 \\
 Adj. $R^2$ &  0.0795 & 0.0184 & 0.181 & 0.0338 & 0.0343 & 0.0572 \\ \hline
\multicolumn{7}{c}{Standardfehler in Klammern} \\
\multicolumn{7}{c}{ *** p$<$0,01, ** p$<$0,05, * p$<$0,1} \\
\end{tabular}}
\end{center}
\end{sidewaystable}

\begin{sidewaystable}

\caption{Ergebnisse der Regressionen f\"ur die Bundesl\"ander Teil 2a}\label{Tab4a}
\begin{center}
\scalebox{0.70}{%
\begin{tabular}{lcccccccccccc} \hline
 & (1) & (2) & (3) & (4) & (5) & (6) \\
VARIABLEN & Relative Ver\"anderung & Relative Ver\"anderung & Relative Ver\"anderung & Relative Ver\"anderung & Relative Ver\"anderung & Relative Ver\"anderung  \\ 
 & Baden-W\"urttemberg & Bayern  & Brandenburg  & Hessen  & Mecklenburg-Vorpommern  & Niedersachsen   \\ \hline
 Methode & -1.398*** & -0.632*** & -1.421*** & -1.173*** & -3.133*** & -1.517***  \\
 & (0.179) & (0.186) & (0.395) & (0.224) & (0.808) & (0.284) &  \\
 Einwohner & -3.70e-06 & 1.51e-06 & 4.13e-06 & -2.91e-06 & 2.14e-05 & -2.24e-06  \\
 & (2.59e-06) & (1.46e-06) & (1.11e-05) & (2.45e-06) & (2.62e-05) & (4.94e-06) \\
Ausl\"anderquote & -3.849** & -8.759*** & 2.925 & -2.793 & -76.87 & 2.295  \\
 & (1.941) & (2.034) & (3.273) & (2.355) & (103.4) & (3.857)  \\
Gender & -18.45*** & -3.425 & 13.91 & -21.59** & 48.52*** & 14.88  \\
 & (6.953) & (3.904) & (12.04) & (10.25) & (12.85) & (9.101) \\
Studentenanteil & -1.843 & 0.683 & -0.539 & -1.874 & 7.222 & -22.86***  \\
 & (2.945) & (3.788) & (4.572) & (1.985) & (12.40) & (6.754)  \\
Constant & 8.581** & 1.292 & -8.059 & 11.02** & -25.43*** & -7.927*  \\
 & (3.479) & (1.949) & (5.985) & (5.155) & (6.428) & (4.589)  \\
Beobachtungen & 1027 & 1905 & 273 & 424 & 236 & 765  \\
$R^2$ & 0.150 & 0.035 & 0.058 & 0.177 & 0.093 & 0.069  \\
 Adj. $R^2$ & 0.146 & 0.0320 & 0.0408 & 0.167 & 0.0733 & 0.0625  \\ \hline
\multicolumn{7}{c}{Standardfehler in Klammern} \\
\multicolumn{7}{c}{ *** p$<$0,01, ** p$<$0,05, * p$<$0,1} \\
\end{tabular}}
\end{center}
\end{sidewaystable}

\begin{sidewaystable}

\caption{Ergebnisse der Regressionen f\"ur die Bundesl\"ander Teil 2b}\label{Tab4b}
\begin{center}
\scalebox{0.70}{%
\begin{tabular}{lcccccccccccc} \hline
 & (7) & (8) & (9) & (10) & (11) & (12) \\
VARIABLEN & Relative Ver\"anderung & Relative Ver\"anderung & Relative Ver\"anderung & Relative Ver\"anderung & Relative Ver\"anderung & Relative Ver\"anderung  \\ 
 & Nordrhein-Westfalen  & Rheinland-Pfalz  & Saarland  & Sachsen  & Schleswig-Holstein  & Th\"uringen    \\ \hline
 Methode & -1.836*** & -1.913*** & -1.210 & -1.116*** & -1.898** & -1.648*** \\
 &  (0.340) & (0.470) & (0.750) & (0.254) & (0.867) & (0.527) \\
 Einwohner &  3.11e-06** & -7.11e-06 & -4.86e-05 & -2.55e-07 & 2.83e-05 & 3.73e-06 \\
 &  (1.50e-06) & (9.39e-06) & (3.69e-05) & (2.48e-06) & (2.00e-05) & (9.32e-06) \\
Ausl\"anderquote &  -4.724 & 11.86*** & 1.622 & -0.675 & -55.77*** & 1.179 \\
 & (3.254) & (3.264) & (7.239) & (1.296) & (17.38) & (11.34) \\
Gender &  4.797 & 35.63*** & -116.8*** & 21.60*** & 21.78 & -19.34** \\
 &  (15.22) & (7.307) & (41.45) & (7.300) & (18.40) & (9.023) \\
Studentenanteil & -19.09*** & -0.699 & 56.59 & -7.930 & -24.96 & 7.406 \\
 & (5.087) & (4.702) & (37.16) & (8.511) & (31.15) & (13.19) \\
Constant &  -2.085 & -18.28*** & 60.27*** & -12.08*** & -9.970 & 8.657* \\
 &  (7.633) & (3.697) & (21.13) & (3.666) & (9.262) & (4.528) \\
Beobachtungen &  396 & 693 & 52 & 457 & 391 & 324 \\
$R^2$ &  0.121 & 0.075 & 0.368 & 0.059 & 0.068 & 0.079 \\
 Adj. $R^2$ &  0.109 & 0.0680 & 0.299 & 0.0485 & 0.0554 & 0.0650 \\ \hline
\multicolumn{7}{c}{Standardfehler in Klammern} \\
\multicolumn{7}{c}{ *** p$<$0,01, ** p$<$0,05, * p$<$0,1} \\
\end{tabular}}
\end{center}
\end{sidewaystable}

\begin{table}
\caption{Ergebnisse der Regressionen mit Melderegisterdaten Teil 1}\label{Tab5a}
\begin{center}
\scalebox{0.80}{%
\begin{tabular}{lccc} \hline
 & (1) & (2) & (3) \\
VARIABLEN & Relative Ver\"anderung & Relative Ver\"anderung & Relative Ver\"anderung \\ 
 & Baden-W\"urttemberg  & Niedersachsen  & Rheinland-Pfalz \\ \hline
Methode & -1.675*** & -1.260*** & -1.675*** \\
 & (0.161) & (0.452) & (0.452) \\
Melderegister & -5.85e-06** & -8.35e-06 & 2.66e-06 \\
 & (2.65e-06) & (6.25e-06) & (8.14e-06) \\
Constant & -0.927*** & -0.195 & 0.350*** \\
 & (0.0726) & (0.327) & (0.108) \\
Beobachtungen & 1,027 & 338 & 463 \\
$R^2$  & 0.134 & 0.039 & 0.043 \\
 Adj. $R^2$  & 0.133 & 0.0329 & 0.0390 \\ \hline
\multicolumn{4}{c}{Standardfehler in Klammern} \\
\multicolumn{4}{c}{ *** p$<$0.01, ** p$<$0.05, * p$<$0.1} \\
\end{tabular}}
\end{center}
\end{table}

\begin{table}
\caption{Ergebnisse der Regressionen mit Melderegisterdaten Teil 2}\label{Tab5b}
\begin{center}
\scalebox{0.80}{%
\begin{tabular}{lccc} \hline
 & (1) & (2) & (3) \\
VARIABLEN & Relative Ver\"anderung & Relative Ver\"anderung & Relative Ver\"anderung \\ 
 &Baden-W\"urttemberg  & Niedersachsen  & Rheinland-Pfalz \\ \hline
 Methode & -1.347*** & -1.235*** & -2.003*** \\
 & (0.181) & (0.473) & (0.453) \\
Melderegister & -2.17e-06 & -4.47e-06 & -5.57e-06 \\
 & (3.04e-06) & (7.62e-06) & (9.30e-06) \\
Ausl\"anderquote & -4.005** & -11.91 & 10.64*** \\
 & (1.945) & (8.555) & (3.338) \\
Gender & -18.76*** & -1.439 & 25.16*** \\
 & (6.959) & (19.71) & (8.067) \\
Studentenanteil & -0.511 & -0.229 & 0.113 \\
 & (0.361) & (1.025) & (0.659) \\
Constant & 8.733** & 0.962 & -12.88*** \\
 & (3.483) & (9.951) & (4.093) \\
Beobachtungen & 1,027 & 338 & 463 \\
$R^2$& 0.148 & 0.045 & 0.085 \\
  Adj. $R^2$ & 0.143 & 0.0302 & 0.0750 \\ \hline
\multicolumn{4}{c}{Standardfehler in Klammern} \\
\multicolumn{4}{c}{ *** p$<$0.01, ** p$<$0.05, * p$<$0.1} \\
\end{tabular}}
\end{center}
\end{table}

\end{appendix}

\end{document}